\documentclass[draft]{agujournal2018}
\draftfalse

\usepackage{booktabs}
\usepackage{url}

\usepackage[bookmarks=false,  pdfnewwindow=true, colorlinks=true,  linkcolor=magenta, citecolor=blue, filecolor=cyan, urlcolor=purple, final=true]{hyperref}

\usepackage{changepage}

\usepackage{journal_macros}

\journalname{Space Weather}

\begin{document}

\sloppy

\title{CME Magnetic Structure and IMF Preconditioning Affecting SEP Transport}

\authors{Erika~Palmerio\affil{1,2,3}, Emilia~K.~J.~Kilpua\affil{1}, Olivier~Witasse\affil{4}, David~Barnes\affil{5}, Beatriz~S{\'a}nchez-Cano\affil{6}, Andreas~J.~Weiss\affil{7,8,9}, Teresa~Nieves-Chinchilla\affil{10}, Christian~{M\"o}stl\affil{7,9}, Lan~K.~Jian\affil{10}, Marilena~Mierla\affil{11,12}, Andrei~N.~Zhukov\affil{11,13}, Jingnan~Guo\affil{14,15}, Luciano~Rodriguez\affil{11}, Patrick~J.~Lowrance\affil{16}, Alexey~Isavnin\affil{17}, Lucile~Turc\affil{1}, Yoshifumi~Futaana\affil{18}, and Mats~Holmstr{\"o}m\affil{18}}

\begin{adjustwidth}{-1cm}{-1cm}
\affiliation{1}{Department of Physics, University of Helsinki, Helsinki, Finland}
\affiliation{2}{Space Sciences Laboratory, University of California--Berkeley, Berkeley, CA, USA}
\affiliation{3}{CPAESS, University Corporation for Atmospheric Research, Boulder, CO, USA}
\affiliation{4}{ESTEC, European Space Agency, Noordwijk, Netherlands}
\affiliation{5}{STFC RAL Space, Rutherford Appleton Laboratory, Harwell Campus, Oxfordshire, UK}
\affiliation{6}{School of Physics and Astronomy, University of Leicester, Leicester, UK}
\affiliation{7}{Space Research Institute, Austrian Academy of Sciences, Graz, Austria}
\affiliation{8}{Institute of Physics, University of Graz, Graz, Austria}
\affiliation{9}{Institute of Geodesy, Graz University of Technology, Graz, Austria}
\affiliation{10}{Heliophysics Science Division, NASA Goddard Space Flight Center, Greenbelt, MD, USA}
\affiliation{11}{Solar--Terrestrial Centre of Excellence---SIDC, Royal Observatory of Belgium, Brussels, Belgium}
\affiliation{12}{Institute of Geodynamics of the Romanian Academy, Bucharest, Romania}
\affiliation{13}{Skobeltsyn Institute of Nuclear Physics, Moscow State University, Moscow, Russia}
\affiliation{14}{School of Earth and Space Sciences, University of Science and Technology of China, Hefei, China}
\affiliation{15}{CAS Center for Excellence in Comparative Planetology, University of Science and Technology of China, Hefei, China}
\affiliation{16}{IPAC--Spitzer, California Institute of Technology, Pasadena, CA, USA}
\affiliation{17}{Rays of Space Oy, Vantaa, Finland}
\affiliation{18}{Swedish Institute of Space Physics, Kiruna, Sweden}
\end{adjustwidth}


\correspondingauthor{Erika Palmerio}{epalmerio@berkeley.edu}

\begin{keypoints}
\item We analyse the 2012~May~11 CME and the 2012~May~17 SEP event through the inner heliosphere
\item The May~11 CME appeared to rotate rapidly upon eruption and then considerably during its interplanetary propagation
\item The May~11 CME provided direct magnetic connectivity for the efficient transport of SEPs from the May~17 event
\end{keypoints}


\begin{abstract}
Coronal mass ejections (CMEs) and solar energetic particles (SEPs) are two phenomena that can cause severe space weather effects throughout the heliosphere. The evolution of CMEs, especially in terms of their magnetic structure, and the configuration of the interplanetary magnetic field (IMF) that influences the transport of SEPs are currently areas of active research. These two aspects are not necessarily independent of each other, especially during solar maximum when multiple eruptive events can occur close in time. Accordingly, we present the analysis of a CME that erupted on 2012 May 11 (SOL2012-05-11) and an SEP event following an eruption that took place on 2012 May 17 (SOL2012-05-17). After observing the May 11 CME using remote-sensing data from three viewpoints, we evaluate its propagation through interplanetary space using several models. Then, we analyse in-situ measurements from five predicted impact locations (Venus, Earth, the Spitzer Space Telescope, the Mars Science Laboratory en route to Mars, and Mars) in order to search for CME signatures. We find that all in-situ locations detect signatures of an SEP event, which we trace back to the May 17 eruption. These findings suggest that the May 11 CME provided a direct magnetic connectivity for the efficient transport of SEPs. We discuss the space weather implications of CME evolution, regarding in particular its magnetic structure, and CME-driven IMF preconditioning that facilitates SEP transport. Finally, this work remarks the importance of using data from multiple spacecraft, even those that do not include space weather research as their primary objective.
\end{abstract}


\section{Introduction} \label{sec:intro}

Solar energetic particle \citep[SEP; e.g.,][]{reames2015,vainio2009} events are increases in high-energy particle fluxes with energies in the keV--GeV range lasting from hours to days and important drivers of space weather effects \citep[e.g.,][]{koskinen2017}. They are intrinsically related to two major classes of eruptions from the Sun, namely flares \citep[e.g.,][]{benz2017} and coronal mass ejections \citep[CMEs; e.g.,][]{webb2012}. These phenomena can take place either together or separately in the solar atmosphere, and both can contribute to the production of SEPs, although in different but yet connected ways. Explosive magnetic reconnection during solar flares can accelerate particles, which then travel along the interplanetary magnetic field \citep[IMF; e.g.,][]{owens2013} lines connecting to an observer. CMEs, on the other hand, consist of copious amounts of plasma and magnetic field that are released into interplanetary space and that can drive shock waves, thus acting as efficient particle accelerators. Contrarily to flares that accelerate particles only on the solar surface, CME-driven shocks may accelerate particles locally in the low corona and also at large distances in the heliosphere, with acceleration sites that typically extend much wider. These two forms of SEP production have traditionally resulted in a clear distinction of particle acceleration processes \citep[e.g.,][]{cane1986,reames2013,vlahos2019} into flare-accelerated \citep[often with an impulsive profile; e.g.,][]{reames1990} and shock-accelerated \citep[more likely to show a gradual time evolution; e.g.,][]{desai2016}. This `dual nature' of SEPs, however, is not fully representative of the complex nature and interplay of processes that result in particle injection and acceleration, given that different mechanisms can contribute to a single event \citep[e.g.,][]{anastasiadis2019,cane2010}.

Since SEPs are accelerated and can propagate more efficiently along magnetic field lines, their spatial distribution is, at least in principle, supposed to be limited to the heliolongitudes (and latitudes) that are magnetically connected to the acceleration site(s) \citep[e.g.,][]{reames1999}. As a result, impulsive SEP events are expected to be observed predominantly within a narrower region compared to gradual ones \citep[e.g.,][]{reames2002,reames2013}. Nevertheless, surprisingly wide (i.e., significantly larger than a flare site or a CME-driven shock front) longitudinal distributions of SEPs have been reported for both impulsive \citep[e.g.,][]{lario2017,wibberenz2006,wiedenbeck2013} and gradual \citep[e.g.,][]{dresing2012,richardson2014a,rouillard2012} events. Possible reasons for wide-spread SEP distributions are cross-field transport in the interplanetary medium and/or an extended source region at the Sun injecting particles over a broad region \citep[e.g.,][]{dresing2014}. Accordingly, from a space weather perspective, current SEP research focuses not only on how intense an event could be, but also on which heliolongitudes it could extend to  \citep[see, e.g., the recent reviews by][]{klein2017,malandraki2018}. 

Being large-scale magnetic disturbances, CMEs profoundly affect the structure of the IMF during their journey away from the Sun \citep[e.g.,][]{witasse2017}. As a result, the passage of a CME may provide a temporary magnetic connection between two regions in the heliosphere that would otherwise not be linked. This may result in energetic particles observed in situ `inside' an interplanetary CME \citep[or ICME; e.g.,][]{kilpua2017b} that preceded the SEP event \citep[e.g.,][]{dresing2016,larson1997,masson2012,rodriguez2008a}. In these efforts, knowledge of the magnetic structure of CMEs in interplanetary space is crucial. Regardless of their pre-eruptive magnetic configuration, it is generally agreed that all CMEs lift off from the Sun as helical magnetic structures called flux ropes \citep[e.g.,][]{forbes2000,green2018,vourlidas2013}, which consist of bundles of magnetic field lines that wind about a common axis. From a space weather perspective, there are several factors to take into account after a CME has left the Sun, i.e.\ its size and propagation direction, which determine whether a CME will impact a certain location \citep[e.g.,][]{mays2015a,mostl2017,rodriguez2011}, its propagation speed, which determines the arrival time \citep[e.g.,][]{verbeke2019a,zhao2014}, and its magnetic structure, which is important in determining the resulting space weather response \citep[e.g.,][]{kilpua2019a,savani2015}. A review summarising the current status of space weather forecasting of CMEs has been recently published by \citet{vourlidas2019}. Whilst hit/miss and arrival time predictions presently lie around a hit rate of ${\sim}0.5$ and an accuracy of ${\pm}10$~hours \citep[e.g.,][]{riley2018,wold2018}, current models can only reproduce, rather than effectively forecast, the magnetic structure of CMEs. This is because, even if the magnetic configuration of flux ropes during eruption can be indirectly estimated from remote-sensing observations \citep[e.g.,][]{gopalswamy2018,palmerio2017}, it can differ significantly when measured in situ \citep[e.g.,][]{palmerio2018,yurchyshyn2008}. Parameters that can influence the evolution of CMEs as they travel away from the Sun are e.g.\ deflections \citep[e.g.,][]{kay2015a,wang2004b,zhuang2017}, rotations \citep[e.g.,][]{isavnin2013,kliem2012,vourlidas2011}, deformations \citep[e.g.,][]{manchester2004,owens2008,savani2010}, and interactions with the ambient solar wind \citep[e.g.,][]{rouillard2010,rodriguez2016, winslow2016}, stream interaction regions \citep[e.g.,][]{heinemann2019,liu2019,prise2015}, and/or with other CMEs \citep[e.g.,][]{kilpua2019b,lugaz2014,scolini2020}. Comprehensive reviews on the interplanetary evolution of CMEs have been recently published by \citet{lugaz2017a}, \citet{luhmann2020}, and \citet{manchester2017}.

In this work, we explore the large-scale preconditioning of the IMF resulting from the passage of an ICME and its subsequent effects on the transport of SEPs. We study the eruption and interplanetary evolution of a CME that erupted on 2012~May~11, with a particular focus on its magnetic structure as it travels away from the Sun. We evaluate the propagation of the CME across the inner heliosphere and how its orientation changed, indicating that the flux rope rotated and was distorted significantly after erupting. Furthermore, we analyse the SEP event following the 2012~May~17 eruption, which was observed in situ at eight well-separated locations in the inner heliosphere, including all four planets therein: Mercury, Venus, Earth and Mars. We show that five of the eight available observers recorded a nearly simultaneous SEP event characterised by an impulsive profile, whilst the remaining three observed a more gradual event. These findings are consistent with the IMF being affected by the passage of an ICME, since impulsive SEP profiles were observed at those locations that were predicted to be encountered by the May~11 CME. We suggest that the ICME provided the required `direct' magnetic connectivity for SEPs to spread rapidly over a broad region (the observing locations engulf over ${\sim}0.9$~AU in radial distance and ${\sim}70^{\circ}$ in longitude, extending up to ${\sim}150^{\circ}$ in longitude away from the flaring site). The 2012~May~17 SEP event was studied extensively in the literature, especially between the Sun and Earth, but, to our knowledge, this is the first report of its observations at eight different locations, five of which were impacted by a preceding ICME.

This article is organised as follows. In Section~\ref{sec:data}, we enumerate the space- and ground-based instruments that are employed in this study. In Section~\ref{sec:remote}, we describe the two eruptive events under analysis (2012~May~11 and May~17) from a remote-sensing observational perspective. In Section~\ref{sec:models}, we estimate the propagation of the May~11 CME and its impact at different locations using several models. In Section~\ref{sec:insitu}, we present the in-situ signatures of the May~11 CME and May~17 SEP event across the inner heliosphere. In Section~\ref{sec:discussion}, we discuss various aspects of the evolution of the May~11 CME in terms of its propagation, its magnetic structure, and its role in the transport of SEPs, by combining observational data with modelling outputs. Finally, in Section~\ref{sec:conclusions}, we conclude by summarising our results.


\section{Spacecraft and Ground-based Data} \label{sec:data}

In this section, we list the fleet of instruments that are involved in this study, in the order in which they are introduced in this article. We use a synthesis of remote-sensing and in-situ data in order to follow and observe signatures of the 2012~May~11 CME and the 2012~May~17 SEP event at different locations throughout the inner heliosphere.

Solar observations from Earth's viewpoint are made with the Solar Dynamics Observatory \citep[SDO;][]{pesnell2012}. The extreme ultra-violet (EUV) images and line-of-sight magnetograms we use are taken with the Atmospheric Imaging Assembly \citep[AIA;][]{lemen2012} and the Helioseismic and Magnetic Imager \citep[HMI;][]{scherrer2012}, respectively. Solar observations from other locations at ${\sim}1$~AU are made with the Sun Earth Connection Coronal and Heliospheric Investigation \citep[SECCHI;][]{howard2008a} Extreme UltraViolet Imager (EUVI) onboard the Solar Terrestrial Relations Observatory \citep[STEREO;][]{kaiser2008}. STEREO consists of twin spacecraft that orbit the Sun, one ahead of Earth in its orbit (STEREO-A) and the other one trailing behind (STEREO-B, which has been out of contact since October 2014). Furthermore, we use data from the X-ray Sensor (XRS) onboard the Geostationary Operational Environmental Satellites (GOES) 15 satellite to study the solar X-ray flux.

Coronagraph observations are also made from three vantage points. The view from Earth is provided by the Large Angle and Spectrometric Coronagraph \citep[LASCO;][]{brueckner1995} C2 (2.2--6 $R_{\odot}$) and C3 (3.5--30 $R_{\odot}$) instruments onboard the Solar and Heliospheric Observatory \citep[SOHO:][]{domingo1995}. The views from STEREO-A and STEREO-B are provided by the SECCHI/COR1 (1.5--4 $R_{\odot}$) and COR2 (2.5 -- 15 $R_{\odot}$) coronagraphs. 

Heliospheric observations are made with the Heliospheric Imagers \citep[HI;][]{eyles2009} onboard the twin STEREO spacecraft. Each HI instrument comprises two cameras, HI1 (4--24$^{\circ}$) and HI2 (18--88$^{\circ}$), that image the space between the Sun and Earth (the degrees measure the elongation in helioprojective radial coordinates).

In-situ measurements around Venus are taken with Venus Express \citep[VEX;][]{svedhem2007}. We use data from the Magnetometer \citep[MAG;][]{zhang2006} and the Analyser of Space Plasmas and Energetic Atoms \citep[ASPERA-4;][]{barabash2007}. The sensors that we use from ASPERA-4 are the Ion Mass Analyser (IMA) and the Electron Spectrometer (ELS), both performing local charged particle measurements. VEX, which ended its operations in 2015, had a 24-hour highly elliptical and quasi-polar orbit, and spent a couple of hours each day inside the induced magnetosphere of Venus. ASPERA-4 was operational for several hours close to periapsis and apoapsis only, whilst the magnetometer ran continuously.

In-situ measurements from near Earth are mainly taken with the Wind satellite, which is operational at Earth's Lagrange L1 point. We use data from the Magnetic Fields Investigation \citep[MFI;][]{lepping1995} and Solar Wind Experiment \citep[SWE;][]{ogilvie1995} instruments, which measure magnetic field and plasma (including solar wind electron distributions) continuously. Additionally, we use proton flux data from the Electron, Proton, and Alpha Detector (EPEAD) instrument onboard GOES-13. We also study variations in cosmic rays and SEPs on the ground using count rate data from the Neutron Monitor Database (NMDB), and in particular from the South Pole (SOPO) neutron monitor.

Measurements around 1~AU are also taken with the Spitzer Space Telescope \citep{werner2004}, which orbits the Sun on an Earth-trailing orbit. In order to evaluate the impact of space weather events at Spitzer, we count the radiation hits on the Infrared Array Camera \citep[IRAC;][]{fazio2004}. Spitzer was deactivated in 2020.

We also analyse data recorded by the Mars Science Laboratory \citep[MSL;][]{grotzinger2012} spacecraft that was en route to Mars at the time of this study. We use data from the Radiation Assessment Detector \citep[RAD;][]{hassler2012} instrument onboard the Curiosity rover, and in particular from the plastic scintillator `E', which measures the radiation dose rate contributed by all particles reaching the detector from all directions and provides the best statistics of background cosmic ray fluxes.

In-situ measurements from Mars are taken with two different spacecraft. The first is Mars Express \citep[MEX;][]{chicarro2004}. We use data from the Mars Advanced Radar for Subsurface and Ionospheric Sounding \citep[MARSIS;][]{picardi2004} and the Analyzer of Space Plasmas and Energetic Atoms \citep[ASPERA-3;][]{barabash2006}. MARSIS is a high-frequency sounding radar dedicated to probe the Martian subsurface, surface, and ionosphere. ASPERA-3 is identical to the ASPERA-4 instrument onboard VEX, and the sensors that we use are again IMA and ELS. MEX has a 7-hour elliptical orbit, with a periapsis distance of ${\sim}300$~km and an apoapsis distance of ${\sim}10000$~km from the planet's surface. MARSIS takes measurements at periapsis only, whilst ASPERA-3 was operational for about half of the orbit during the time of the events under study. The second satellite is 2001 Mars Odyssey \citep[MOdy;][]{saunders2004}, from which we use data taken with the High Energy Neutron Detector (HEND), which is part of the Gamma Ray Spectrometer \citep[GRS;][]{boynton2004} suite.

Finally, we inspect SEP measurements made at the twin STEREO spacecraft and Mercury. At STEREO, we use data from the High Energy Telescope \citep[HET;][]{vonrosenvinge2008}, which is part of the In situ Measurements of Particles And CME Transients \citep[IMPACT;][]{luhmann2008} investigation. At Mercury, we use data from the Mercury Surface, Space Environment, Geochemistry, and Ranging \citep[MESSENGER;][]{solomon2007} spacecraft, which orbited the innermost planet between 2011 and 2015. We examine measurements taken with the Neutron Spectrometer (NS) sensor, part of the Gamma-Ray and Neutron Spectrometer \citep[GRNS;][]{goldsten2007} instrument and dedicated to measuring the flux of ejected neutrons.


\section{Remote-sensing Observations} \label{sec:remote}

\begin{figure}[ht]
\centering
\includegraphics[width=.9\linewidth]{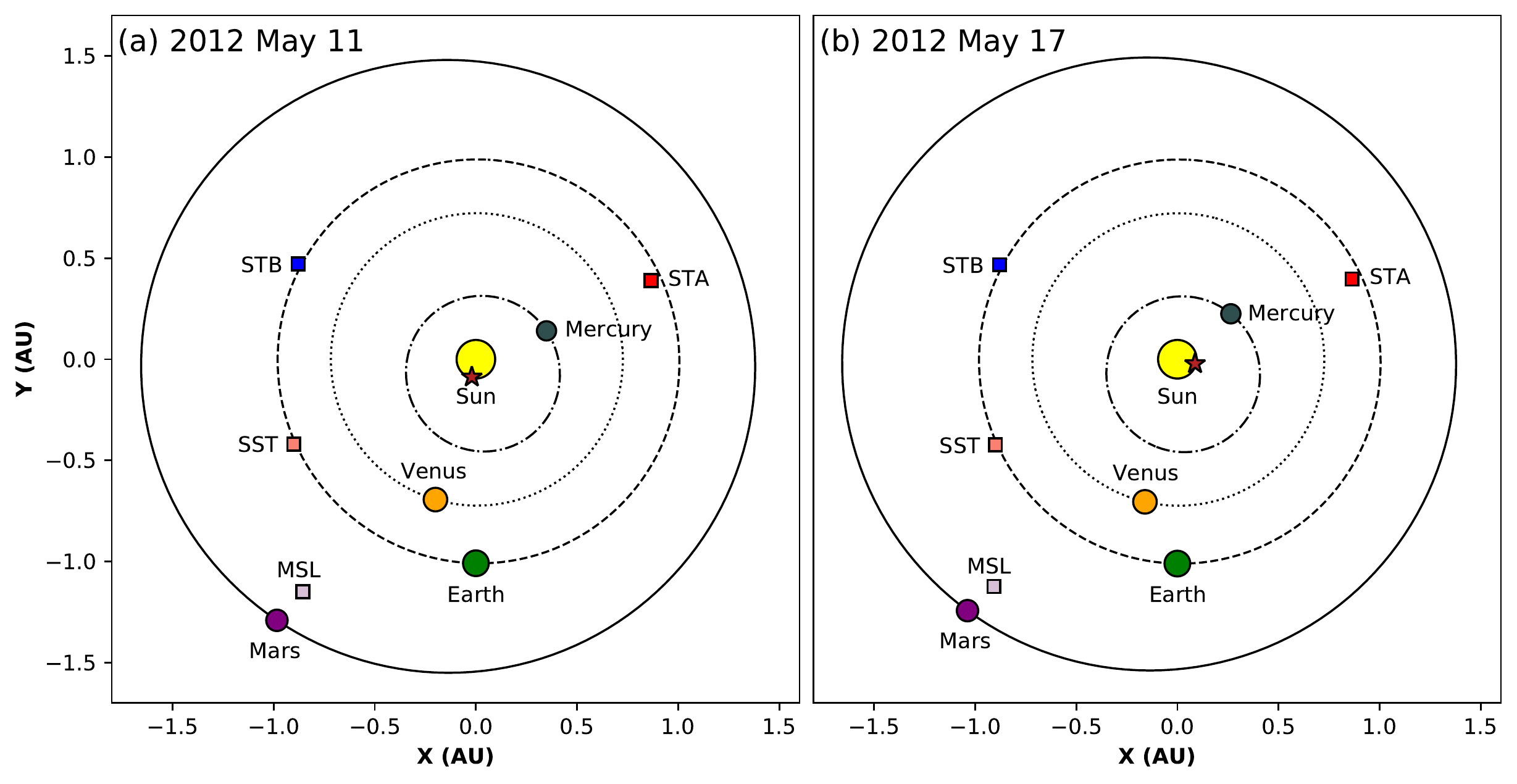}
\caption{Position of various planets and spacecraft in the inner solar system (i.e., up to Mars' orbit) on (a) 2012~May~11 and (b) 2012~May~17, projected onto the solar equatorial plane. The projected source locations of the May~11 and May~17 eruptions are indicated by a star symbol on the surface of the Sun in panels (a) and (b), respectively. The planets are marked with circles, whilst the spacecraft are marked with squares (STA = STEREO-A; STB = STEREO-B; SST = Spitzer Space Telescope; MSL = Mars Science Laboratory). The orbits of all planets are also indicated.}
\label{fig:map}
\end{figure}

In this section, we describe the eruptive events of 2012 May~11 (Section~\ref{subsec:may11}) and May~17 (Section~\ref{subsec:may17}) from a remote-sensing observational perspective. The positions of the inner planets and the spacecraft employed in this study, together with the eruptions' source region locations at the Sun, are shown in Figure~\ref{fig:map} for (a) May~11 and (b) May~17.

\subsection{The 2012 May 11 Eruption} \label{subsec:may11}

The first eruptive event that we focus on in this work initiated from a small active region containing a filament on 2012~May~11 around 23:00~UT. In this section, we follow the eruption and subsequent propagation of the associated large-scale CME using remote-sensing observations of the solar disc, the solar corona, and the heliosphere.

\subsubsection{Solar Disc Observations} \label{subsubsec:disc}

\begin{figure}[ht]
\centering
\includegraphics[width=.99\linewidth]{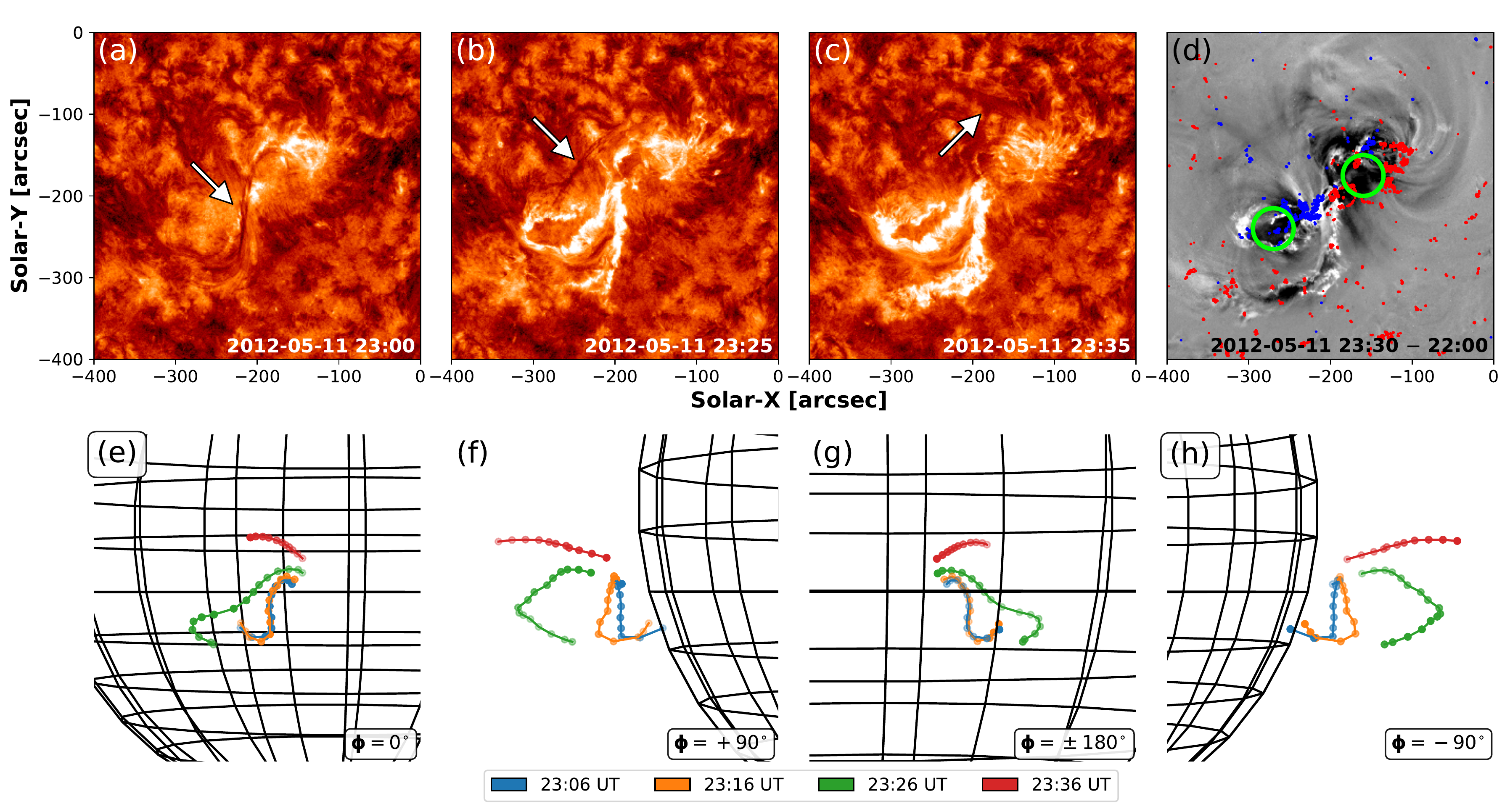}
\caption{The 2012~May~11 eruption from solar disc data. (a--c) The erupting filament as seen by SDO/AIA in the 304~{\AA} channel, with the rotating filament indicated by an arrow in each panel. (d) Base-difference image taken with SDO/AIA in the 211~{\AA} channel and overlaid with SDO/HMI magnetogram contours (blue = negative polarity, red = positive polarity). The dimming regions (signatures of the flux rope footpoints) have been circled in green. (e--h) The erupting filament triangulated with the tie-pointing technique using SDO/AIA and STEREO/SECCHI/EUVI-B data in the 304~{\AA} channel. The same reconstruction is shown from four different longitudinal perspectives \citep[indicated in each panel in Stonyhurst coordinates;][]{thompson2006} to facilitate its 3D visualisation. Each colour represents one different timestamp within the same panel.}
\label{fig:eruption}
\end{figure}

The active region from which the 2012~May~11 CME originated appeared on the eastern limb of the Earth-facing Sun on 2012~May~6, on the southern hemisphere. It was not attributed a National Oceanic and Atmospheric Administration (NOAA) classification number, but it had Space-weather HMI Active Region Patch \citep[SHARP;][]{bobra2014} number 1642. An S-shaped filament was clearly visible within the active region throughout its rotation towards the central meridian. The filament erupted on 2012~May~11, around 23:00~UT, from approximately S13E13. The top panels of Figure~\ref{fig:eruption} show the evolution of the eruption process in remote-sensing data from SDO's perspective. For the complete set of observations of the erupting filament from three viewpoints (STEREO-A, SDO, and STEREO-B), see Movie~S1.

The magnetic structure of the flux rope associated with the erupting CME, or intrinsic flux rope type, can be estimated using a combination of several indirect proxies at different wavelengths in remote-sensing data of the solar disc \citep[for a summary of the proxies that can be used to determine chirality, tilt, and field direction at the axis of a flux rope, see][]{palmerio2017}. In the particular case under study, we note that the forward S-shape of the erupting filament (Figure~\ref{fig:eruption}a) is a sign of right-handed chirality \citep[e.g.,][]{martin2012}. Furthermore, the forward J-shape of the flare ribbons associated with the eruption (bright features in Figure~\ref{fig:eruption}b--c) confirms the right-handedness of the flux rope \citep[e.g.,][]{demoulin1996}. We infer the tilt of the flux rope axis by taking the average inclination between the polarity inversion line (PIL) and post-eruption arcades (PEAs). We determine the tilt of the flux rope to be ${\sim}65^{\circ}$ with respect to the solar equator. Finally, we determine the magnetic field direction at the flux rope axis from coronal dimmings, which are believed to map to the CME footpoints \citep[e.g.,][]{thompson2000}. The dimming regions marked in Figure~\ref{fig:eruption}d indicate that the axial field was directed roughly towards the south at the time of the eruption. Thus, we determine the intrinsic flux rope type of the CME under study to be right-handed with a mostly-southward axial field, or east--south--west (ESW) according to the convention of \citet{bothmer1998} and \citet{mulligan1998}. 

Upon eruption, the filament spine was seen to rotate clearly clockwise (Figure \ref{fig:eruption}a--c), as expected for a right-handed flux rope \citep[e.g.,][]{fan2004,green2007,lynch2009}. Its southern leg (i.e., the eastern leg upon rotation) could no longer be seen in SDO images shortly after the eruption, whilst its northern leg (i.e., the western leg upon rotation) could be observed for a couple of hours after the eruption onset. STEREO-B observations (see Movie~S1), taken from an almost-quadrature view with respect to SDO, show that the southeastern leg of the filament appeared to disconnect from the Sun approximately at the time of its disappearance in SDO imagery. Such asymmetric filament eruptions have been observed in previous studies \citep[e.g.,][]{liu2009,tripathi2006}. According to the definition of \citet{liu2009}, the filament eruption studied here can be classified as whipping-like, where the filament ``active'' leg whips upwards and the ``anchored'' leg remains fixed to the photosphere. In such a scenario, the mass in the active leg could either fall back towards the Sun or fail to follow the motion of the filament spine, thus showing an apparent detachment of the leg from the solar surface, to which the filament field may however still be connected. Another possibility is that the active leg undergoes interchange reconnection with a nearby coronal hole open field \citep[e.g.,][]{baker2009,zhu2014}, hence opening one end of the filament to interplanetary space. In the event studied here, the motion of the flare ribbon brightenings to the south of the CME source region (visible in Figure~\ref{fig:eruption}b--c and in Movie~S1), corresponding to the extent of a small region of open field (visible in SDO/AIA 193~{\AA} imagery, data not shown), supports this latter scenario.

The bottom panels of Figure~\ref{fig:eruption} show 3D reconstructions of the filament during the early phase of its eruption using the tie-pointing triangulation technique. The method was first employed by \citet{thompson2009} to triangulate a Sun-grazing comet, but it has been used since then also to evaluate the 3D rotation of erupting filaments \citep[e.g.,][]{bemporad2011,thompson2012}. The reconstructions presented here are obtained using SDO/AIA and STEREO/SECCHI/EUVI-B data, both in the 304~{\AA} channel. At the fourth and last triangulation (23:36~UT), the southeastern (active) leg of the filament had completely disappeared, hence we could only triangulate the northwestern (anchored) one. It should be emphasised that erupting filaments are thick in EUV images and the resulting triangulated structures are expected to be accompanied by significant uncertainties. Consequently, the resulting structures shown in Figure~\ref{fig:eruption}e--h should be considered only as approximate indicators of the global 3D shape of the erupting filament. Nevertheless, regardless of the connection of the active filament leg to the Sun, the observations reported here suggest that the filament changed its orientation from a roughly perpendicular to a roughly parallel one with respect to the solar equator, rotating by ${\sim}85^{\circ}$ clockwise. Whether the associated flux rope experienced the same evolution, however, is unclear, as the spatial relationship between erupting filaments and the overlying flux rope is an open question \citep[e.g.,][]{gibson2006a,vourlidas2013}. If we assume that the orientation of a flux rope follows that of its corresponding filament, we can deduce that the flux rope type changed from ESW to north--east--south (NES) due to such rotation.

\subsubsection{Coronagraph Observations} \label{subsubsec:coronagraph}

After erupting, the 2012~May~11 CME was visible in coronagraphs from three well-separated viewpoints (SOHO, STEREO-A, and STEREO-B) as it propagated through the solar corona (see Movie~S2). The CME is seen pointing towards the west in STEREO-B and towards the east in STEREO-A imagery, indicating an Earth-directed eruption. However, we note from SOHO images that the CME propagated slightly eastwards rather that along the Sun--Earth line, indicating that the eruption was directed roughly towards Mars (see the positions of different planets and spacecraft in Figure~\ref{fig:map}a).

\begin{figure}[ht]
\centering
\includegraphics[width=.99\linewidth]{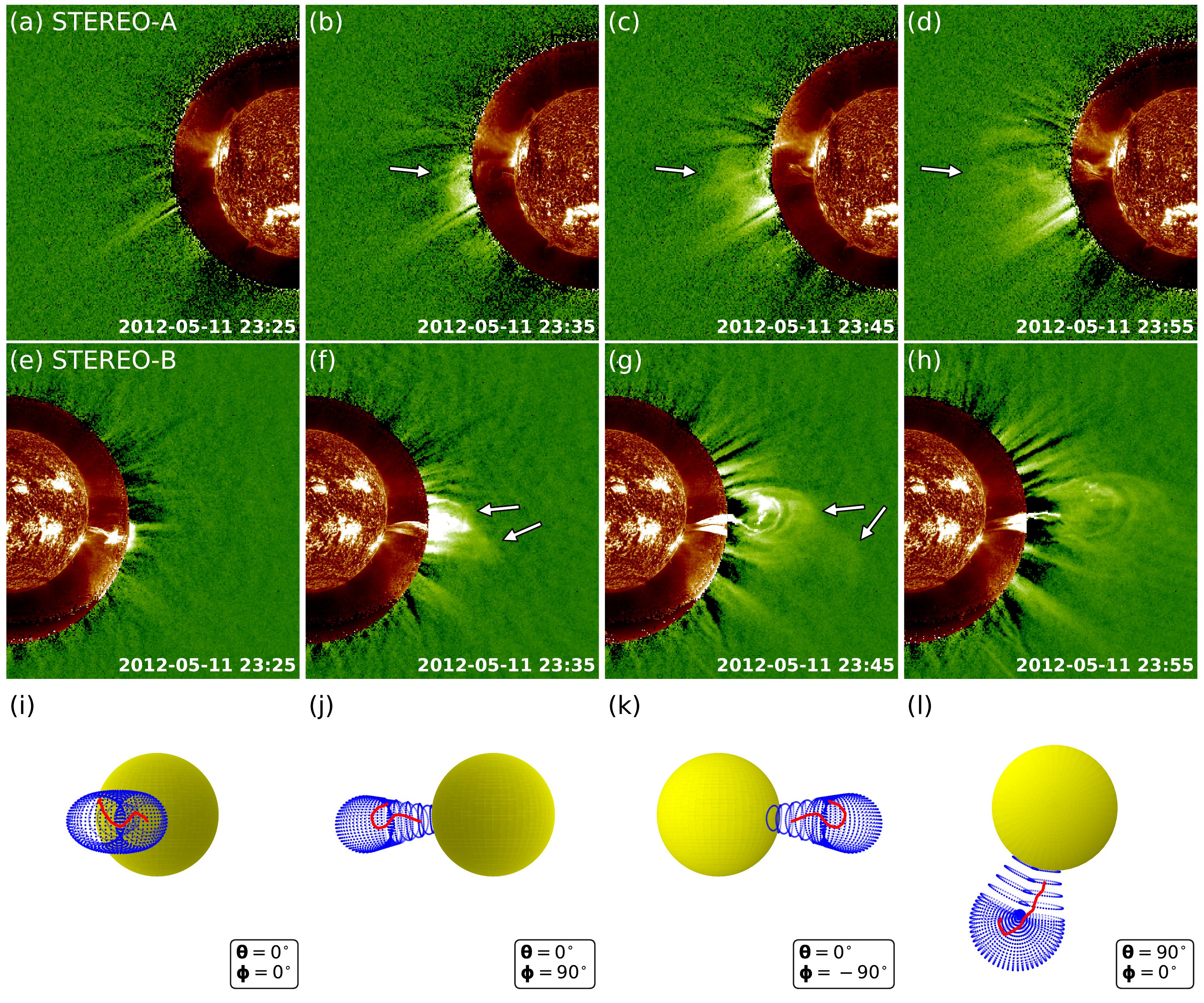}
\caption{(a--h) Evolution of the erupting filament and the overlying CME (marked by arrows in some of the panels) through the low corona in composite images from STEREO/SECCHI/EUVI in the 304~{\AA} channel and STEREO/SECCHI/COR1 using (a--d) STEREO-A and (e--h) STEREO-B data. The COR1 data are shown as base-difference images, with the background taken at 23:00~UT. (i--l) 3D reconstruction of the erupting filament and the overlying large-scale CME performed at 23:45~UT. The filament (shown in red) is triangulated with the tie-pointing technique using STEREO/SECCHI/EUVI in the 304~{\AA} channel and COR1 composite data from both STEREO spacecraft. The overlying CME (shown as a blue wireframe) is reconstructed with the GCS technique using STEREO/SECCHI/COR1 data from both STEREO spacecraft. The same reconstruction is shown from different perspectives (indicated in each panel in Stonyhurst coordinates) to facilitate its 3D visualisation.}
\label{fig:cor1}
\end{figure}

The top two rows of Figure~\ref{fig:cor1} show composite images using EUVI and COR1 data from both STEREO-A (panels a--d) and STEREO-B (panels e--h). In particular, STEREO-B imagery shows that the filament leg that disappeared first as seen from SDO (i.e., the active leg, Figure~\ref{fig:eruption}a--c) rotated significantly from a downward-facing hook (Figure~\ref{fig:cor1}e) to an upward-facing hook (Figure~\ref{fig:cor1}g) morphology. The western leg appeared to be still attached to the Sun throughout the rotation phase. The corresponding large-scale CME, in turn, featured a single-front structure from the STEREO-A perspective (marked by an arrow in Figure~\ref{fig:cor1}b--d and visible in Movie~S2) and a double-bump structured front from the STEREO-B and SOHO viewpoints (marked by arrows in Figure~\ref{fig:cor1}f--g and visible in Movie~S2). We note that asymmetric white-light CMEs associated with whipping-like filament eruptions were reported in previous studies, e.g.\ by \citet{zhu2014}.

The bottom row of Figure~\ref{fig:cor1} shows 3D reconstructions of the filament, performed using the tie-pointing technique on composite EUVI and COR1 images from both STEREO spacecraft, and the overlying large-scale CME, performed using the graduated cylindrical shell \citep[GCS;][]{thernisien2006,thernisien2009} reconstruction technique on COR1 images from both STEREO spacecraft. The morphology of the GCS model is that of a hollow wireframe, with six free parameters that can be manually adjusted until they best match the data. For the reconstruction shown here, we only include the ``main'' structure seen in STEREO-B (i.e., that indicated by the top arrow in Figure~\ref{fig:cor1}g). According to the resulting 3D images shown in Figure~\ref{fig:cor1}i--l, the western leg of the filament (i.e., the anchored leg discussed in Section~\ref{subsubsec:disc}) follows the western leg of the larger-scale CME quite closely, and the detachment and slight rotation northwards of the active filament leg are also evident. To the thick nature of filaments discussed in Section~\ref{subsubsec:disc}, we add that in this case the tie-pointing reconstruction was applied to simultaneous EUV and white-light images, i.e.\ using instruments that measure at different wavelengths, resulting in large uncertainties. Nevertheless, considering how well the resulting 3D thread fits within the larger CME structure, these results can be considered a good approximation of the general morphology and relative orientation of the two features. The tilt of the CME inferred from the GCS reconstruction is in this case close to $0^{\circ}$, meaning that the eruption was seen basically edge-on from both STEREO spacecraft. This also implies that it is not possible to determine whether the eastern CME leg, which corresponds to the active filament leg, was still attached to the Sun at this point, either entirely or at least partially. Nevertheless, the low tilt angle inferred from the GCS reconstruction is consistent with the filament rotation (see Section~\ref{subsubsec:disc} and Figure~\ref{fig:eruption}), suggesting that the flux rope axis roughly followed the orientation of the underlying filament and had a NES orientation in the low corona.

\begin{figure}[ht]
\centering
\includegraphics[width=.99\linewidth]{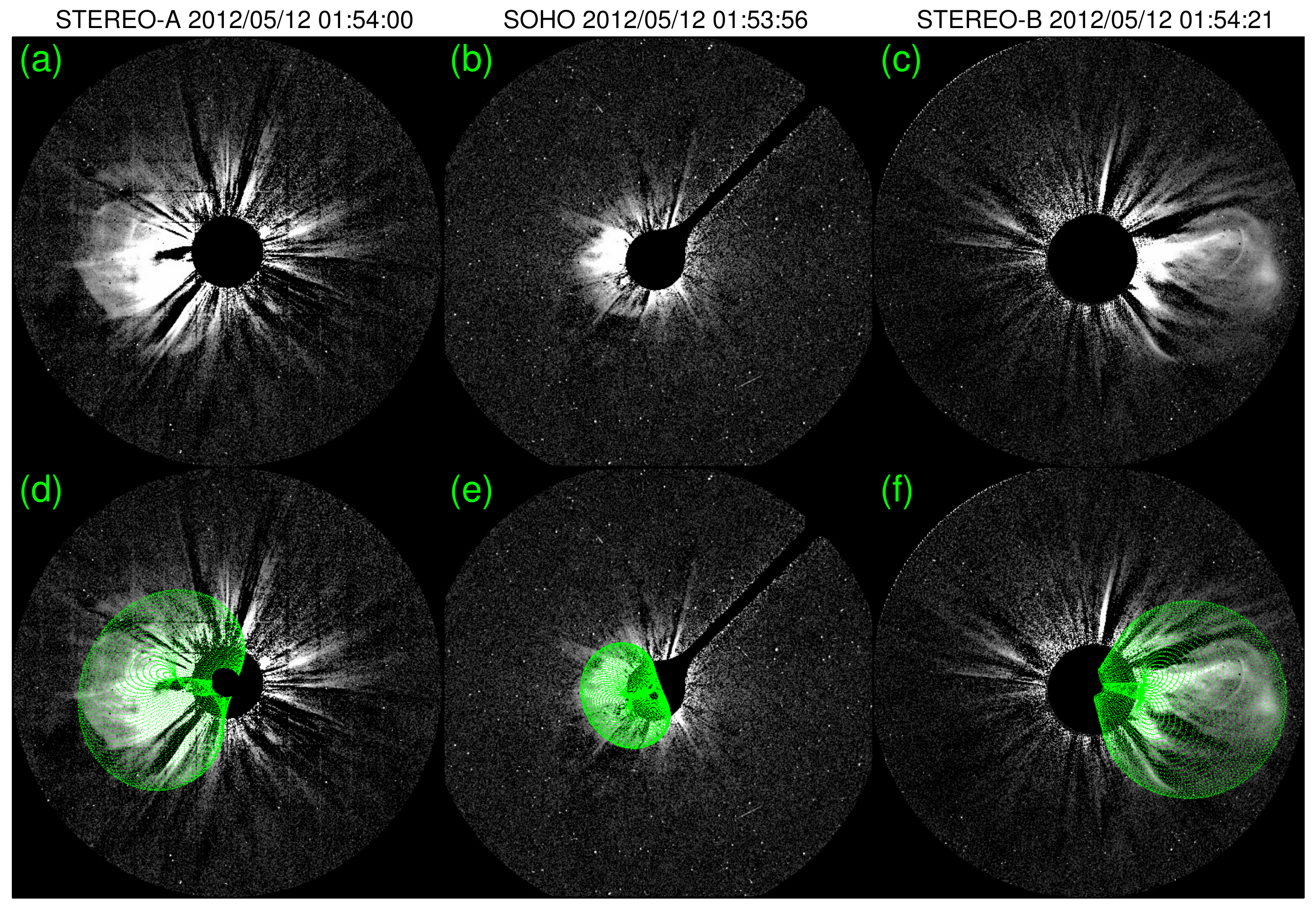}
\caption{Example of the GCS reconstruction technique applied to the 2012~May~11 CME at the last available time before the CME left the COR2-B field of view (2012~May~12, 01:54~UT). (a--c) Base-difference images taken from STEREO/SECCHI/COR2-A, SOHO/LASCO/C3, and STEREO/SECCHI/COR2-B. (d--f) Same images as in (a--c), with the GCS wireframe overlaid.}
\label{fig:gcs}
\end{figure}

Finally, in order to define the global geometric and kinematic parameters of the 2012~May~11 CME during its journey through the solar corona, we further apply the GCS technique to coronagraph images from all three available spacecraft (i.e., SOHO and the twin STEREO). An example of the resulting GCS reconstructions is shown in Figure~\ref{fig:gcs} (for the full CME kinematics, see Figure~S1). Throughout its passage through the coronagraphic fields of view, the CME did not seem to experience significant deflections, featuring a propagation direction of $-10^{\circ}$ in latitude and $-30^{\circ}$ in longitude in Stonyhurst coordinates. On the other hand, the CME seemed to continue rotating as it travelled away from the Sun, with its axis inclination resulting from our reconstructions evolving gradually from $-10^{\circ}$ to $-65^{\circ}$ (a positive tilt value is defined for counterclockwise rotations) with respect to the solar equatorial plane. Finally, its propagation speed was ${\sim}1000$~km$\cdot$s$^{-1}$ (calculated between successive reconstructions from the CME apex height) throughout the COR2 field of view. However, consisting of a ``hollow croissant'', the GCS model cannot provide information on the internal magnetic field. Hence, although the handedness of the CME is known from solar disc observations (see Section~\ref{subsubsec:disc}), the direction of the corresponding flux rope's axial field is characterised by a $180^{\circ}$ ambiguity. If we assume that the flux rope followed the same rotation pattern as its associated filament, then it can be concluded that the flux rope axis rotated ${\sim}130^{\circ}$ from its pre-eruptive configuration to altitudes of a couple tens solar radii. According to these assumptions and minding that the CME was right-handed, it would follow that the flux rope left the outer corona close to a west--north--east (WNE) type, whilst a $180^{\circ}$ reversal of the axis would yield a east--south--west (ESW) type. We remark that, as shown in Figure~\ref{fig:cor1}, Figure~\ref{fig:gcs}, and Movie~S2, the CME under study appeared highly asymmetric, thus the GCS results should be considered as an approximation of the global orientation of a significantly distorted structure. Furthermore, we note that for this event it is particularly difficult to clearly discern the boundaries between the CME itself and the white-light shock driven by it. Hence, being the 2012~May~11 CME far from a ``textbook'' case, we expect GCS reconstructions to not be able to unequivocally derive the 3D structure of the CME in white light.

\subsubsection{HI Observations} \label{subsubsec:hi}

\begin{figure}[ht]
\centering
\includegraphics[width=.99\linewidth]{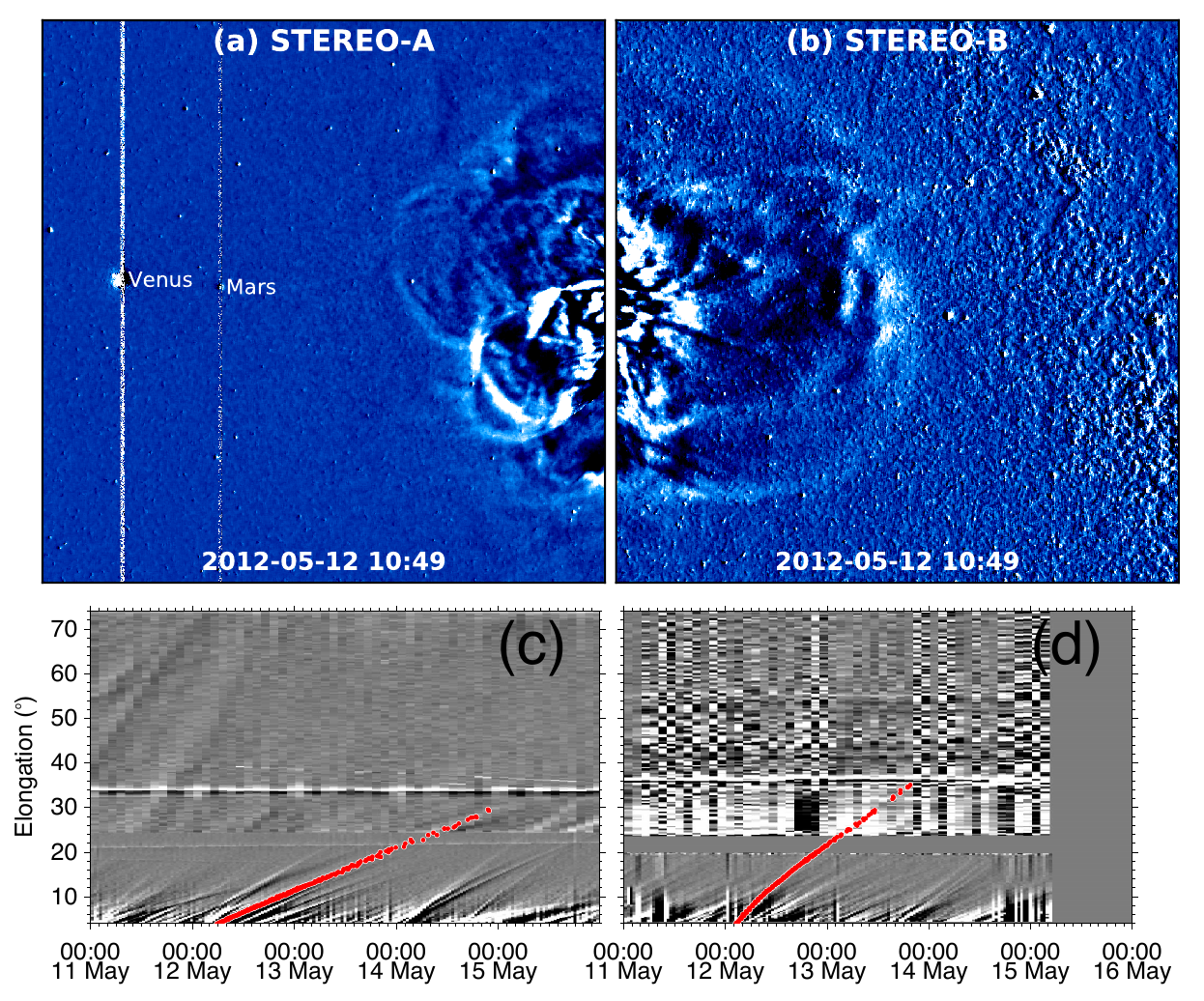}
\caption{The 2012~May~11 CME as seen in images from the STEREO/SECCHI/HI cameras. (a--b) Snapshots of the CME in running-difference images taken with the (a) HI1-A and (b) HI1-B cameras. The projected locations of Venus and Mars are marked in panel (a). (c--d) Time--elongation maps from (c) STEREO-A and (d) STEREO-B. The CME tracks are marked in red. The maps are constructed along position angles of $90^{\circ}$ for STEREO-A and $264^{\circ}$ for STEREO-B, i.e.\ close to the ecliptic plane (the position angle of Earth was ${\sim}89.9^{\circ}$ from the STEREO-A perspective and ${\sim}264.4^{\circ}$ from the STEREO-B perspective).}
\label{fig:hi}
\end{figure}

After leaving the STEREO/SECCHI/COR2 field of view, the 2012~May~11 CME appeared in the HI cameras onboard both STEREO spacecraft (see Movie~S3). Figure~\ref{fig:hi}a--b shows a snapshot of the CME as seen in the HI1 field of view. At both spacecraft, observations of the CME in HI1 began at approximately 02:00~UT on 2012~May~12. The event was clearly visible from both STEREO viewpoints, and the CME apex was observed at progressively larger elongation angles in STEREO-B imagery compared to STEREO-A (note the position of the apex in Figure~\ref{fig:hi} and in Movie~S3). This suggests that the CME was closer to quadrature-view from the STEREO-B perspective, consistently with the coronagraph analysis reported in Section~\ref{subsubsec:coronagraph} (see Figure~\ref{fig:map}a for the position of the STEREO spacecraft).

In order to follow the propagation of the CME under study through both HI cameras, we use time--elongation maps \citep[e.g.][]{sheeley2008,davies2009} produced from running-difference images. Within these time--elongation maps, a propagating structure such as a CME appears as a bright front followed by a dark front. This is due to the increase and subsequent decrease in density and allows features to be tracked in elongation as a function of time. This is performed on the CME under investigation here, which we track for over 2.5~days in HI-A to an elongation of ${\sim}30^\circ$ and for over 1.5~days in HI-B to an elongation of ${\sim}35^{\circ}$. The resulting time--elongation maps, with the corresponding CME tracks, are shown in Figure~\ref{fig:hi}c--d.

Furthermore, the CME under study is listed in the HELiospheric Cataloguing, Analysis and Techniques Service (HELCATS) catalogues. The HELCATS project ran from 2014 to 2017 and aimed, amongst other goals, to catalogue and analyse solar transients (such as CMEs) detected in the STEREO/SECCHI/HI cameras. This event is included in the HICAT catalogue \citep{harrison2018}, which was generated through visual inspection of background-subtracted and difference HI1 images, and in the HIGeoCAT catalogue \citep{barnes2019}, which was generated using time--elongation maps and applying single-spacecraft fitting techniques to derive CME kinematic properties. In both catalogues, the CME is labelled as \textsc{HCME\_A\_\kern-.14ex\_20120511\_01} for STEREO-A and \textsc{HCME\_B\_\kern-.14ex\_20120512\_01} for STEREO-B. We remark that, in both HICAT and HIGeoCAT, CMEs are identified using single-spacecraft data, hence the STEREO-A and STEREO-B observations are reported separately. Of the fitting techniques used in HIGeoCAT, we report here the results obtained with the Self-Similar Expansion \citep[SSE;][]{davies2012,mostl2013} fitting technique with a fixed half-width of $30^{\circ}$ (hereafter, SSEF30) applied to time--elongation single-spacecraft data. In the SSE model, CMEs are assumed to have a circular front and to propagate radially with a constant speed and half-width. We note that the SSEF30 results obtained using STEREO-B data are consistent with the GCS results reported in Section~\ref{subsubsec:coronagraph}, i.e., the CME propagates in direction $(\theta,\phi) = (-9^{\circ},-19^{\circ})$ and with a speed of $869$~km$\cdot$s$^{-1}$. SSEF30 results based on STEREO-A data, however, are significantly different, reporting a propagation direction of $(\theta,\phi) = (-4^{\circ},-54^{\circ})$ and a speed of $2008$~km$\cdot$s$^{-1}$. Since, as stated above, the 2012~May~11 CME was closer to quadrature view (i.e., with less projection effects) from STEREO-B than from STEREO-A, we expect the fitting results retrieved from STEREO-B data to be more accurate.

\subsection{The 2012 May 17 Eruption} \label{subsec:may17}

\begin{figure}[ht]
\centering
\includegraphics[width=.99\linewidth]{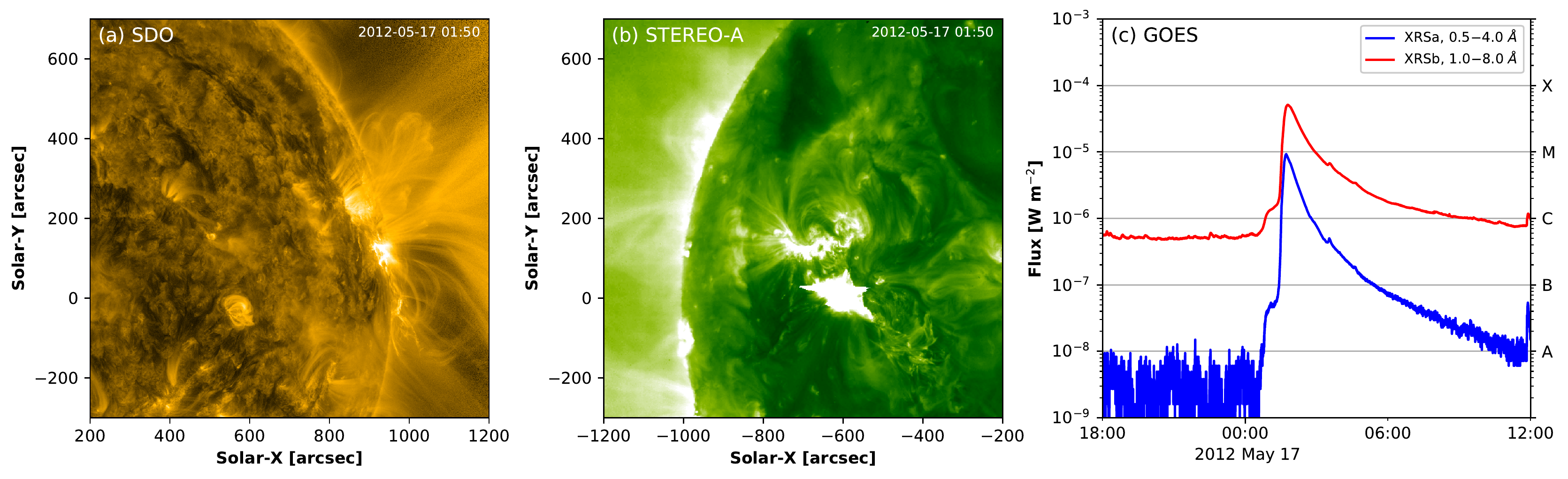}
\caption{Overview of the 2012~May~17 eruption. (a) SDO/AIA image of the source region (AR 11476) in the 171~{\AA} channel shortly after the flare onset time. (b) STEREO/SECCHI/EUVI-A image of the source region in the 195~{\AA} channel taken at the same time as (a). (c) GOES-15/XRS soft X-ray flux, showing the occurrence of an M5.1 flare.}
\label{fig:flare}
\end{figure}

The second eruptive event that we focus on in this work initiated from AR 11476 on 2012~May~17 around 01:00~UT. Since in this case we are mostly interested in the release of energetic particles, rather than the CME eruption and evolution itself, we provide here a brief overview of the event. The source region was located close to the western limb from Earth's perspective (N11W76) and on disc (on the north-eastern quadrant) from STEREO-A's perspective, whilst it was fully back-sided from STEREO-B's viewpoint (see Figure~\ref{fig:map}b). The large-scale CME associated with this eruption was fast, with a speed above 1500~km$\cdot$s$^{-1}$, and large, as it appeared as a halo from all three viewpoints. An intense ``snowstorm'', caused by high-energy protons striking the cameras, was seen in both SOHO/LASCO coronagraphs (C2 and C3) starting around 02:00~UT. Furthermore, the GOES/XRS instrument reported an M5.1 flare associated with this event, starting at 01:25~UT, peaking at 01:47~UT, and ending at 02:14~UT. Figure~\ref{fig:flare} provides images of the CME's source region as seen by SDO (Figure~\ref{fig:flare}a) and STEREO-A (Figure~\ref{fig:flare}b), together with the soft X-ray flux measured by GOES-15 (Figure~\ref{fig:flare}c).

The 2012~May~17 eruption has been studied, e.g., by \citet{gopalswamy2013}, \citet{li2013}, \citet{shen2013}, and \citet{rouillard2016}, hence the reader is referred to these articles for additional information and images on the eruptive event and the CME's propagation through the solar corona. In particular, we note that \citet{gopalswamy2013} estimated the shock formation and SEP release heights for the May~17 CME to be 1.38\,$R_{\odot}$ (at 01:32~UT) and 2.32\,$R_{\odot}$ (at 01:40~UT) from the solar centre, respectively. Together with the occurrence of the M5.1 flare, this suggests that the 2012~May~17 SEP event had contributions from both flare-accelerated and shock-accelerated particles, which is what \citet{cane2010} concluded to be the most likely scenario for large SEP events. This conclusion was, in fact, reached by \citet{li2013}, who estimated that electrons were accelerated by the flare at 01:29~UT and protons were accelerated by the CME-driven shock at 01:39~UT from an altitude of 3.07\,$R_{\odot}$. Furthermore, \citet{shen2013} suggested that the 2012~May~17 CME was composed of two nearly-simultaneous ejecta that erupted from the same active region and neutral line, leading to a CME--CME shock interaction scenario. Finally, previous studies reported also the presence of a so-called EUV wave \citep[e.g.,][]{thompson1998,zhukov2004}, visible from both SDO's and STEREO-A's viewpoints.

\section{CME Propagation Modelling} \label{sec:models}

In this section, we propagate the 2012~May~11 CME using different techniques and evaluate its impact at different planets and spacecraft scattered throughout the inner heliosphere. The results of the propagation models that we consider in this work are summarised in Table~\ref{tab:propag}.

\begin{table}[ht]
\caption{CME arrival times from the different modelling techniques presented in Section~\ref{sec:models}.}
\label{tab:propag}
\center
\begin{tabular}{l @{\hskip .3in} c @{\hskip .1in} c @{\hskip .1in} c @{\hskip .1in} c @{\hskip .1in} c}
\toprule
Model & Venus & Earth & Spitzer & MSL & Mars \\
\midrule
SSEF30--A & --- & --- & 05/12 20:18 & 05/13 07:55 & 05/13 11:51\\
SSEF30--B & 05/13 09:11 & 05/14 04:40 & --- & 05/15 03:18 & 05/15 13:52\\
SSSE &  05/14 19:15 & 05/17 18:51 & 05/15 15:01 & 05/17 00:13 & 05/17 14:57\\
DBM & 05/14 00:08 & 05/15 00:20 & 05/14 23:42 & 05/16 14:42 & 05/17 05:49 \\
Enlil (S)& 05/13 19:33 & 05/14 21:30 & 05/14 20:03 & 05/16 05:54 & 05/16 21:58\\
Enlil (E)& 05/14 01:00 & 05/15 17:52 & 05/15 21:07 & 05/17 03:06 & 05/17 23:27\\
\bottomrule
\end{tabular}
\begin{tablenotes}
\textit{Note.} Dates are shown in the format MM/DD HH:MM. The notations `(S)' and `(E)' denote shock and ejecta, respectively.
\end{tablenotes}
\end{table}

\subsection{(S)SSE Propagation} \label{subsec:ssse}

We first evaluate the impact of the 2012~May~11 CME at various locations in interplanetary space using HI data (see Section~\ref{subsubsec:hi}). Considering again the HELCATS products, we initially search for the CME under study in the ARRCAT catalogue \citep{mostl2017}, which was generated from the list of events in HIGeoCAT by predicting their impact throughout the heliosphere. The predictions were made using the SSEF30 model introduced in Section~\ref{subsubsec:hi}. We note that, in ARRCAT, \textsc{HCME\_A\_\kern-.14ex\_20120511\_01} is reported to arrive at MSL and Mars, whilst \textsc{HCME\_B\_\kern-.14ex\_20120512\_01} is predicted to impact Venus and Earth as well. The exact arrival times predicted by ARRCAT (SSEF30-A and SSEF30-B) are reported in Table~\ref{tab:propag}. We emphasise that Spitzer is not included as a posible target in ARRCAT. Nevertheless, we evaluate arrival estimates at the spacecraft using the CME parameters listed in HIGeoCAT and the equations reported in \citet{mostl2013}. This results in an impact according to SSEF30-A and a close miss according to SSEF30-B.

Since the CME under study was well-visible in both STEREO spacecraft, we also use the two-spacecraft version of the SSE model, i.e.\ the Stereoscopic Self-Similar Expansion \citep[SSSE;][]{davies2013} model, to triangulate the CME position over the ${\sim}1.5$~days in which the event was observed by both spacecraft using time--elongation data. This model also assumes that the CME possesses a circular cross-section in the plane in which the CME is observed (in our case, this is the ecliptic) and a constant half-width. To estimate the half-width to use in this case, we use the formulas in \citet{rodriguez2011} to calculate the maximum angular extent of a CME in both latitude and longitude using the GCS parameters as input. This results in an half-angular extent of $58^{\circ}$ in latitude and $54^{\circ}$ in longitude. Since the SSSE is a 2D model, we are only interested in the longitudinal extent of the CME, hence we set a half-width of $54^{\circ}$. In order to extrapolate the position of the CME beyond the time it was last observed, we fit a second-order polynomial to the distance of the CME apex as a function of time. We also assume that the CME continues to propagate in a constant direction beyond its last observed value. As a result of this extrapolation we expect the CME front to pass over Venus, Earth, Spitzer, MSL, and Mars. Table~\ref{tab:propag} reports the arrival times at all the impacted locations, and Figure~S2 shows the position of the tracked CME front together with the resulting CME arrival times and speeds at the three planets (Venus, Earth, and Mars).

\subsection{DBM Propagation} \label{subsec:dbm}

As a further indication of the CME propagation and impact at different planets and spacecraft, we the drag-based model \citep[DBM;][]{vrsnak2013}, which computes analytically the propagation of CMEs using aerodynamic drag equations and with the assumption of a constant background speed and constant drag parameter. For our run, we use an ambient solar wind speed of $450$~km$\cdot$s$^{-1}$ (from measurements of the solar wind speed at the spacecraft in the inner heliosphere a few days after the eruption time) and a drag parameter of $1{\times}10^{-7}$~km$^{-1}$ \citep[i.e., the mean value found by][]{vrsnak2013}. In the simplest version of the DBM, the geometry of CMEs is that of a 2D circular arc centred at the Sun and moving outwards. The CME parameters that we introduce in the tool are entirely derived from the GCS reconstructions presented in Section~\ref{subsubsec:coronagraph}: height of $14.4$\,$R_{\odot}$ on 2012~May~12 at 01:54~UT, speed of $1005$~km$\cdot$s$^{-1}$ (which is the speed between the reconstructions at 01:24 and 01:54~UT, when the CME reached the height of $14.4$\,$R_{\odot}$), half-width of $54^{\circ}$ \citep[again, the full longitudinal extent of the CME using the formulas by][]{rodriguez2011}, and longitude of $-30^{\circ}$ (in Stonyhurst coordinates). With the initial set of parameters described above, impacts are estimated at Venus, Earth, Spitzer, MSL, and Mars. The resulting arrival times at the different planets and spacecraft are reported in Table~\ref{tab:propag}, and a visual representation of the CME propagated with the DBM is shown in Figure~S3.

\subsection{Enlil Simulation} \label{subsec:enlil}

\begin{figure}[ht]
\centering
\includegraphics[width=.9\linewidth]{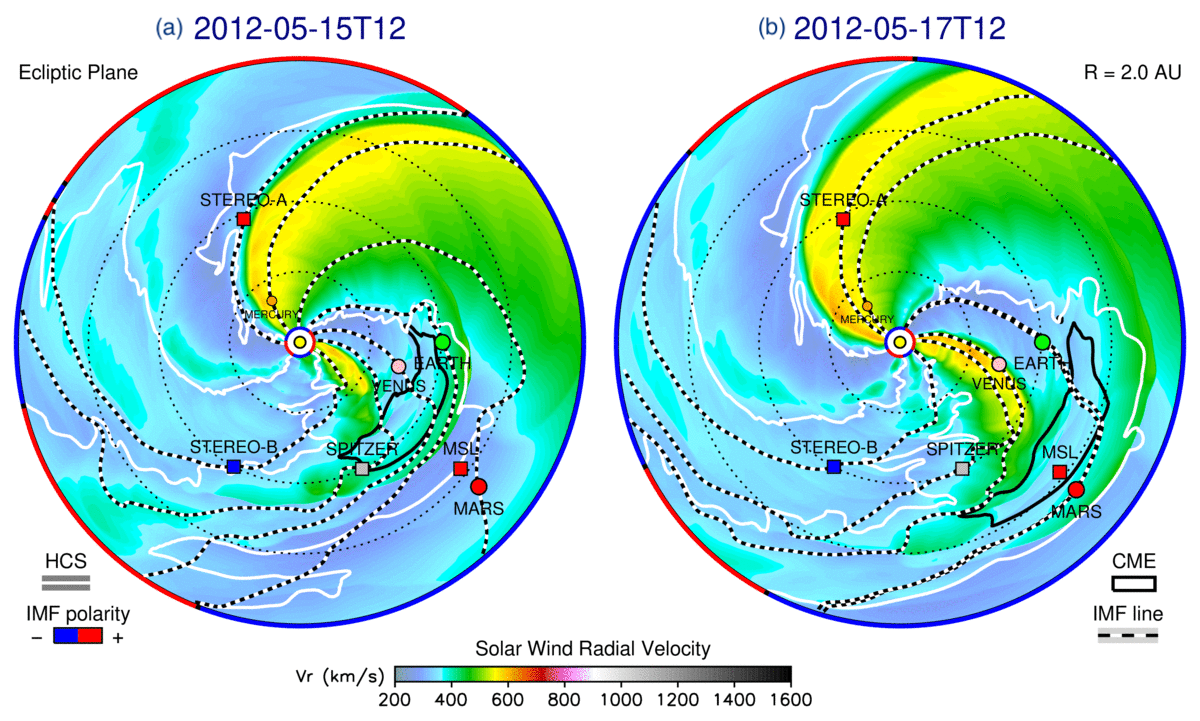}
\caption{Screenshots from the WSA--Enlil+Cone simulation. The parameter shown in the plots is the solar wind radial speed in the ecliptic plane on (a) 2012~May~15, 12:00~UT, and (b) 2012~May~17, 12:00~UT.}
\label{fig:enlil}
\end{figure}

The final CME propagation model that we employ in this study is the 3D heliospheric magnetohydrodynamic (MHD) Enlil \citep{odstrcil2003,odstrcil2004} model. Enlil uses the Wang--Sheeley--Arge \citep[WSA;][]{arge2004} coronal model to simulate the background solar wind from its inner boundary (located at $21.5$\,$R_{\odot}$ or 0.1~AU) onwards. In this case, we set the outer boundary of the simulation domain at 2~AU, i.e.\ including the whole inner heliosphere. CMEs can be modelled through insertion at the inner boundary of the heliospheric domain. In this work, we employ the WSA--Enlil+Cone model, in which CMEs are injected as spherical hydrodynamic structures that lack an internal magnetic field (i.e., there is no internal flux rope structure). Again, we derive the CME initial parameters from GCS reconstructions (see Section~\ref{subsubsec:coronagraph}). We obtain the injection time at the inner boundary by propagating the CME from its last GCS reconstruction (on 2012~May~12 at 01:54~UT) up to $21.5$\,$R_{\odot}$ using a constant speed derived from this last GCS reconstruction and a reconstruction made from data obtained 30~minutes earlier (in this case, at 01:24~UT). This yields an injection time of 2012~May~12 at 03:16~UT with a speed of 1005~km$\cdot$s$^{-1}$. The CME cone that we model has an elliptical cross-section, and we derive its dimensions by `cutting' an elliptical cross-section out of the GCS shell \citep[based on][]{thernisien2011}. The resulting values for the half-angular extent of the major and minor radii are $46.75^{\circ}$ and $37.89^{\circ}$, respectively. Finally, the values for latitude ($-10^{\circ}$), longitude ($-30^{\circ}$), and tilt angle ($-65^{\circ}$) are taken directly from GCS results. The resulting arrival times at different locations throughout the heliosphere are reported in Table~\ref{tab:propag}, and two screenshots from the simulation results are shown in Figure~\ref{fig:enlil}. The two separate rows for Enlil reported in Table~\ref{tab:propag} refer to the shock (S) and ejecta (E) arrival times. As was the case for the previous models, we expect the CME to impact Venus, Earth, Spitzer, MSL, and Mars.

\section{In-situ Measurements} \label{sec:insitu}

Next, we analyse in-situ data from multiple locations scattered throughout the inner heliosphere to evaluate the predicted impacts presented in Section~\ref{sec:models}. Namely, we search for interplanetary signatures of the 2012~May~11 CME at Venus (0.7~AU), Earth (1.0~AU), Spitzer (1.0~AU), MSL (1.4~AU), and Mars (1.6~AU). At each location, in addition to looking for ICME signatures from the 2012~May~11 CME, we search for SEP signatures from the 2012~May~17 event.

\subsection{Measurements at Venus} \label{subsec:venus}

\begin{figure}[ht!]
\centering
\includegraphics[width=.6\linewidth]{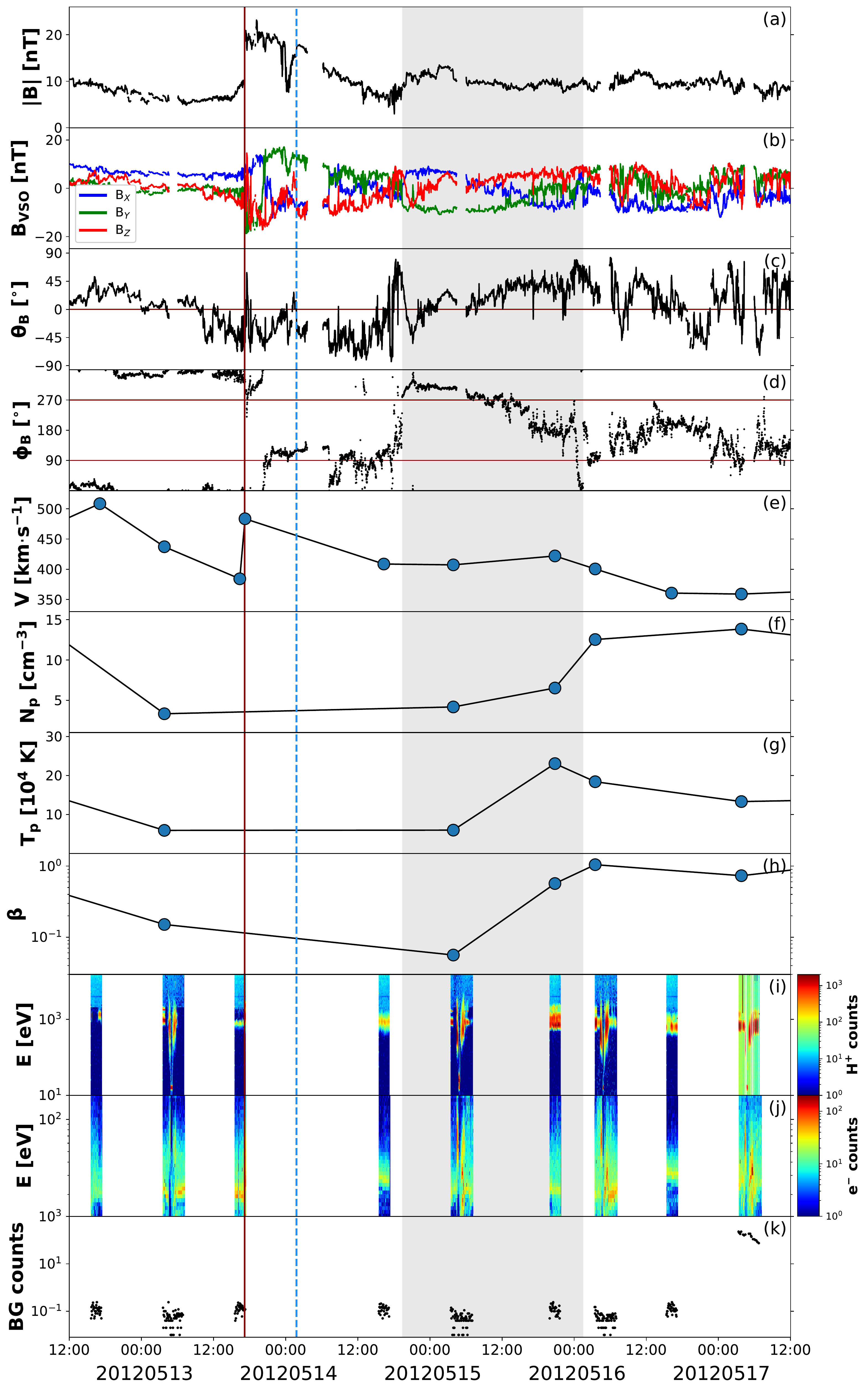}
\caption{Measurements at Venus around the expected arrival time of the 2012~May~11 CME, revealing the passage of an interplanetary disturbance. All data are taken from VEX. The solid vertical line indicates the arrival of the interplanetary shock, whilst the dashed vertical line marks a possible interface between the preceding material and the following sheath driven by the 2012~May~11 CME. The shaded area corresponds to the estimated flux rope-like interval. The periods in which VEX is within Venus' bow shock have been removed from the magnetic field dataset. The parameters shown are: (a) magnetic field magnitude, (b) magnetic field components in Venus Solar Orbital (VSO) Cartesian coordinates, (c) $\theta$ and (d) $\phi$ angles of the magnetic field in VSO angular coordinates, (e) solar wind speed, (f) proton number density, (g) proton temperature, (h) plasma $\beta$, (i) proton and (j) electron energy distribution, and (k) background counts.}
\label{fig:venus}
\end{figure}

The first (in terms of distance from the Sun, see Figure~\ref{fig:map}) location for which an impact of the 2012~May~11 CME is predicted is Venus. Indeed, observations around Venus made by the VEX spacecraft following the eruption reveal a period of transient disturbances. Such measurements are reported in Figure~\ref{fig:venus}. In particular, an interplanetary shock was detected by VEX on 2012~May~13 at 17:10~UT. After a long-duration sheath region, flux rope-like signatures could be identified from 2012~May~14 at 19:23~UT through 2012~May~16 at 01:30~UT. We utilise here the terminology `flux rope-like' rather than the more common `magnetic cloud' because the boundaries were determined from magnetic field data only, since plasma data are not provided by VEX continuously (as explained in Section~\ref{sec:data}, the ASPERA-4 instrument was operational at periapsis and apoaxis only, i.e. about twice per Earth day). The long duration of the sheath region between the interplanetary shock and the following ejecta is likely the result of the interaction of the May~11 CME with a small preceding interplanetary structure. Solar observations from STEREO prior to the eruption of the May~11 CME reveal the presence of several minor eruptions characterised by a jet-like morphology in coronagraph images \citep[see][for a classification of CME morphology types]{vourlidas2013,vourlidas2017} that were possibly Earth-directed. Observations following the eruption of the May~11 CME did not feature CME events large enough to cause the interplanetary signatures shown in Figure~\ref{fig:venus}, indicating that the Sun--Venus connection of the CME is likely correct.

A possible interpretation for the features observed before the flux rope-like ejecta is that the interplanetary shock driven by the 2012~May~11 CME (solid vertical line in Figure~\ref{fig:venus}) propagated through the preceding structure, whilst the following sheath material remained behind. Previous studies have shown that a faster shock can indeed travel through a slower, preceding structure \citep[e.g.,][]{kilpua2019b,lugaz2013}. A possible `interface' between the preceding interplanetary structure and the following sheath region driven by the 2012~May~11 CME is indicated by a dashed vertical line in Figure~\ref{fig:venus} (estimated via the rapid change in direction in the magnetic field $Z$-component). The speed profile within the flux rope-like structure (shaded region in Figure~\ref{fig:venus}) appears nearly flat, suggesting that the ejecta was not expanding as it passed Venus. We emphasise, however, that only two velocity data points fall within the flux rope-like structure and, thus, these conclusions may not be representative of the fine structure of the speed profile. Visual inspection of the ejecta magnetic field shows a rotation of the $Y$-component from west to east and a rotation of the $Z$-component from south to north. This suggests that the corresponding flux rope is right-handed and at an intermediate orientation between a south--west--north (SWN) and a west--north--east (WNE) type. We also estimate the orientation of the flux rope using two techniques. The first is the minimum variance analysis \citep[MVA;][]{sonnerup1967}, where the flux rope axis corresponds to the MVA intermediate variance direction. The orientation of the flux rope axis resulting from MVA is ($\Theta$, $\Phi$) = ($50^{\circ}$, $261^{\circ}$), thus in the intermediate state between a SWN- and a WNE-type flux rope, consistently with what is suggested by visual inspection of the magnetic field data. We also fit the flux rope using the analytical Circular--Cylindrical (CC) model described in \citet{nieveschinchilla2016}. According to the CC model, the flux rope is right-handed, its axis has orientation ($\Theta$, $\Phi$) = ($24^{\circ}$, $256^{\circ}$), and the impact parameter is $y_{0}/R=-0.22$, with $R = 0.15$~AU being the radius of the cloud. The two methods yield an almost identical $\Phi$ angle for the flux rope axis, whilst the $\Theta$ angle differs by ${\sim}25^{\circ}$. Nevertheless, given the usual uncertainties related to all fitting techniques \citep[e.g.,][]{demoulin2018,lepping2003,riley2004}, these results can be deemed mostly consistent, thus indicating a right-handed flux rope with a low-to-intermediate inclination.

Finally, we note a remarkable increase in the background counts measured by VEX/IMA (Figure~\ref{fig:venus}k) during the early hours of 2012~May~17, indicating that an SEP event had impacted Venus. The enhancement of background levels in the ASPERA suite, in fact, corresponds to sufficiently energetic particles that are able to penetrate the instrument \citep[e.g.,][]{futaana2008,ramstad2018}. Since ASPERA-4 was operational close to periapsis and apoapsis only, the background count enhancement was first observed at 03:22~UT after a data gap, thus it is not possible to establish the `true' onset time and peak intensity of the SEP event at Venus. According to the May~17 eruption overview presented in Section~\ref{subsec:may17} and taking into account a particle propagation time of ${\sim}10$--$15$~minutes, we would expect various locations in the inner heliosphere to see an SEP event some time before 02:00~UT on 2012~May~17. This is consistent with the `descending' profile seen in the IMA background counts, suggesting an earlier onset and peak. Thus, the background enhancement seen at VEX seems to be indeed due to the May~17 eruption. Furthermore, we note that the SEP event at Venus is not observed inside the 2012~May~11 ICME ejecta, but ${\sim}1$~day after the passage of what we defined as the trailing edge of the flux rope-like structure. However, considering the CME propagation direction estimated in Sections~\ref{subsec:may11} and \ref{sec:models}, it is likely that Venus was immersed in the CME leg, CME wake, or some other trailing structure in the IMF at the time of the SEP event.

\subsection{Measurements at Earth} \label{subsec:earth}

\begin{figure}[ht!]
\centering
\includegraphics[width=.6\linewidth]{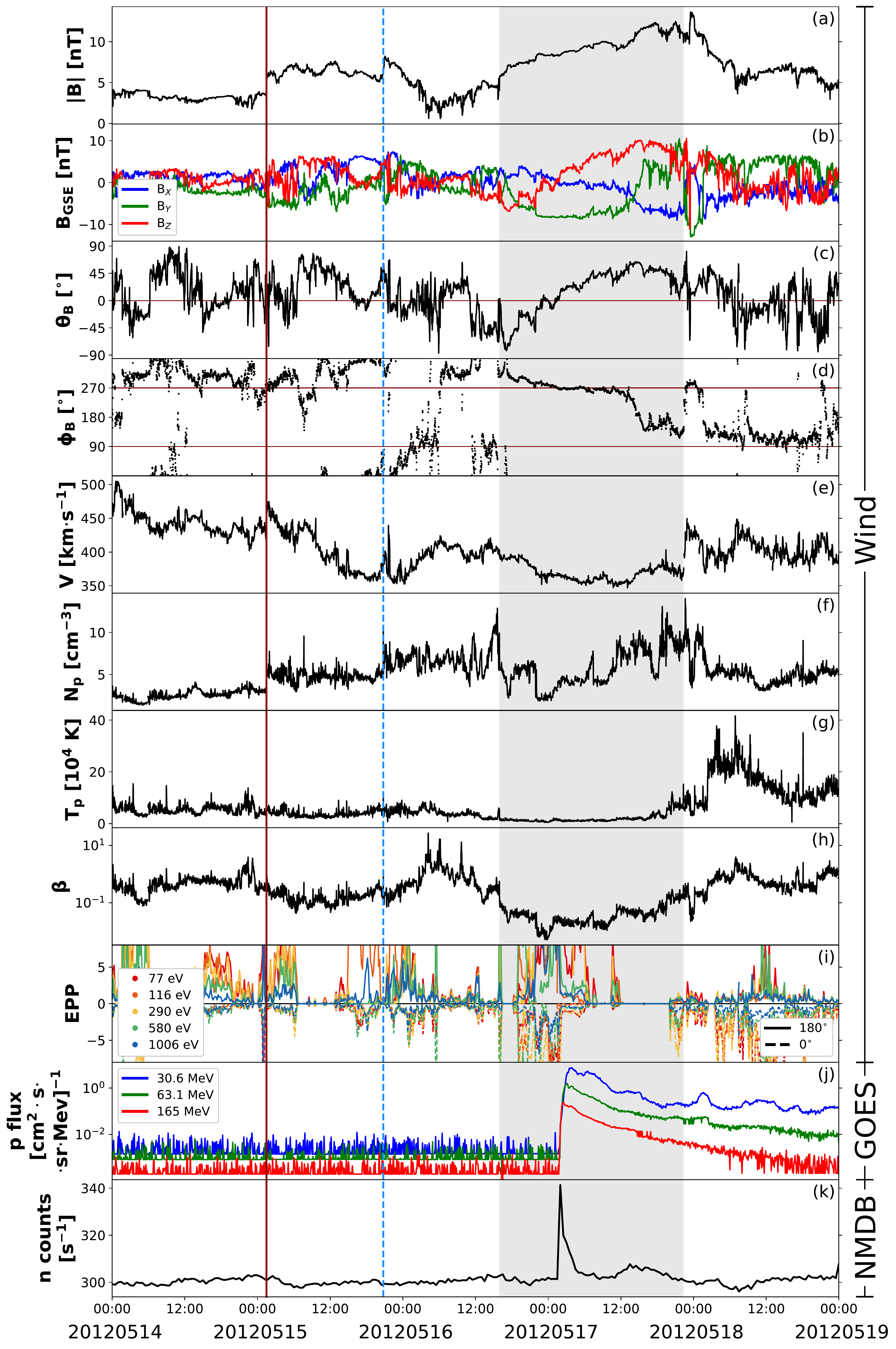}
\caption{Measurements at Earth around the expected arrival time of the 2012~May~11 CME, revealing the passage of an interplanetary disturbance. Data are taken from (a--i) Wind, (j) GOES-13, and (k) the SOPO neutron monitor. The solid vertical line indicates the arrival of the interplanetary shock, whilst the dashed vertical line marks a possible interface between the preceding material and the following sheath driven by the 2012~May~11 CME. The shaded area corresponds to the estimated magnetic cloud interval. The parameters shown are: (a) magnetic field magnitude, (b) magnetic field components in Geocentric Solar Ecliptic (GSE) Cartesian coordinates, (c) $\theta$ and (d) $\phi$ angles of the magnetic field in GSE angular coordinates, (e) solar wind speed, (f) proton number density, (g) proton temperature, (h) plasma $\beta$, (i) electron pitch angle distribution parameter for five energy levels, (j) energetic proton flux, and (k) neutron monitor counts per second.}
\label{fig:earth}
\end{figure}

The next location for which we evaluate a possible impact of the 2012~May~11 CME and 2012~May~17 SEP event is Earth. Given the relatively small separation between Venus and Earth (${\sim}0.3$~AU in radial distance and ${\sim}15^{\circ}$ in longitude, see Figure~\ref{fig:map}), we expect to observe relatively similar ICME signatures that, with the aid of continuous plasma measurements, could strengthen the interpretation reported in Section~\ref{subsec:venus}. Indeed, similarly to observations around Venus, measurements of the solar wind taken from Earth's Lagrange L1 point during the days following the 2012~May~11 eruption reveal clear signatures of a transient period of disturbed IMF and plasma flow. Figure~\ref{fig:earth} shows magnetic field and plasma data taken by Wind together with GOES proton flux and neutron monitor data from the SOPO station. Such measurements show that an interplanetary shock impacted on 2012~May~15 at 01:25~UT, followed by a long-duration sheath. Clear magnetic cloud signatures were visible from 2012~May~16 at 15:57~UT through 2012~May~17 at 22:20~UT.

The sequence of features within the ICME generally matches the measurements at Venus (Section~\ref{subsec:venus}), the interplanetary shock (solid vertical line in Figure~\ref{fig:earth}) being followed by material that appears to belong to two interacting structures. Again, we have marked in Figure~\ref{fig:earth} (with a dashed vertical line) the possible `interface' separating the two. Interestingly, the first portion shows significant expansion from Venus to Earth (from 8.6~hours to 19.3~hours), whilst the following portion of sheath features only minimal expansion (from 17.6~hours to 19.2~hours). This can be explained considering the ambient solar wind ahead of the ICME: at Venus (Figure~\ref{fig:venus}e), the interplanetary shock is preceded by the trailing portion of a fast stream and the following slow wind, whilst at Earth (Figure~\ref{fig:earth}e), the shock has travelled through the slow stream and has caught up with the fast one. This has likely facilitated the expansion of the small structure immediately following the shock, whilst the remaining portion of material (which we associated to the sheath region of the 2012~May~11 CME) shows signatures of compression (with an increasing speed profile and enhanced density) rather than expansion.

The magnetic cloud-type ICME ejecta (shaded region in Figure~\ref{fig:earth}) is reported in both the Richardson~\&~Cane ICME list \citep[hereafter R{\&}C list;][]{cane2003,richardson2010} and the NASA--Wind ICME list \citep[hereafter N-C list;][]{nieveschinchilla2018,nieveschinchilla2019}. The R{\&}C list reports the ejecta measured at Earth as a clear magnetic cloud, featuring bidirectional electrons and lacking signatures of expansion. We also note that, in the R{\&}C list, the 2012~May~11 CME that we analysed through remote-sensing imaging in Section~\ref{sec:remote} is reported as the most probable solar counterpart of this event. The N-C list reports the ejecta as a flux rope, with an apparent expansion velocity of $-7$~km~s$^{-1}$ and a distortion parameter of 0.57 (the distortion parameter is defined as the fraction of the magnetic obstacle where 50\% of the total magnetic field magnitude is accumulated). Again, these signatures show no expansion of the ejecta, but rather its slight compression at the back, which is consistent with the presence of faster wind following it. Indeed, the magnetic cloud hardly features any expansion even when considering its evolution from Venus to Earth (its duration goes from 30.1~hours at Venus to 30.4~hours at Earth). Figure~\ref{fig:earth}i shows the electron pitch-angle-distribution (PAD) parameter (EPP) for five energy levels. The EPP was defined by \citet{nieveschinchilla2016} and consists of the average electron intensities close to $0^{\circ}$ and $180^{\circ}$ normalised to those close to $90^{\circ}$, thus quantifying the bidirectionality of electrons within a magnetic obstacle from PAD data. The profiles show signatures of bidirectionality during the first half of the flux rope and a drop to near zero for the second half. This suggests that the magnetic field lines at the front are still connected to the Sun, whilst in the rear part at least one leg appears to be disconnected (PAD spectra exhibit signatures of one-directional strahl, data not shown). Regarding the magnetic structure of the magnetic cloud, visual inspection of the magnetic field shows a rotation of the $Y$-component from west to east and a rotation of the $Z$-component from south to north. The orientation of the flux rope axis from MVA is ($\Theta$, $\Phi$) = ($46^{\circ}$, $271^{\circ}$), which is almost identical to the orientation found at Venus using the same method. The N-C list also provides fitting results for the flux rope using the CC model, according to which the flux rope is right-handed, its axis has orientation ($\Theta$, $\Phi$) = ($29^{\circ}$, $229^{\circ}$), and the impact parameter is $y_{0}/R=-0.38$, with $R = 0.13$~AU being the radius of the cloud. In addition to the uncertainties related to flux rope fitting techniques mentioned in Section~\ref{subsec:venus}, in this case the difference in axis orientation from the MVA and CC methods may also depend on the fact that the flux rope boundaries do not coincide exactly (in the N-C list, the trailing edge is marked about 4~hours later than the one considered in this work). The importance of the boundary selection to increase the level of agreement across different models was quantified by \citet{alhaddad2013}. Nevertheless, both results are consistent with a right-handed flux rope with a low-to-intermediate inclination (i.e., somewhere between a SWN- and a WNE-type), in agreement with measurements at Venus (Section~\ref{subsec:venus}).

Finally, Figure~\ref{fig:earth}j--k shows proton flux and neutron monitor data. A Forbush decrease \citep{forbush1937,hess1937} can be seen in the SOPO time series starting around 02:30~UT on 2012~May~15, i.e.\ shortly after the interplanetary shock arrival at Wind. We remark that neutron monitor data are collected from ground-based stations, whilst Wind is located at Earth's L1 point, thus the slight (${\sim}1$-hour) delay indicates that the features clearly correspond to the same event. This Forbush decrease was also analysed by \citet{freiherrvonforstner2019}, who connected 45 ICMEs from the Sun to the MSL spacecraft using STEREO/HI data (accordingly, more information regarding the findings of this study can be found in Section~\ref{subsec:msl}). The most striking feature in the neutron monitor time series, however, is the considerable peak in the count rate starting at 01:56~UT on 2012~May~17, i.e.\ during the passage of the magnetic cloud at Earth, indicating the occurrence of a ground-level enhancement \citep[GLE; e.g.,][]{nitta2012}. This corresponds to the SEP event registered by GOES in proton flux measurements, with onset between 01:55 and 02:00~UT. The 2012~May~17 SEP event at Earth was studied in detail in the literature \citep[e.g.,][]{battarbee2018,ding2016,gopalswamy2013,li2013,plainaki2014,rouillard2016}, being associated with the first GLE of solar cycle 24 (GLE71), and was even detected at the International Space Station \citep{berrilli2014}. \citet{rouillard2016} suggested that the passage of the magnetic cloud facilitated the magnetic connectivity between the May~17 eruption source region and Earth.

\subsection{Measurements at the Spitzer Space Telescope} \label{subsec:spitzer}

Also at ${\sim}1$~AU but separated by ${\sim}70^{\circ}$ from Earth (see Figure~\ref{fig:map}) was Spitzer. Although this telescope was dedicated to infrared measurements of deep space, it has been shown that it can be used to investigate space weather events \citep{cheng2014}. Specifically, energetic particles of both solar and extra-solar origin can impact multiple subsystems on the spacecraft, hence Spitzer can act as a space weather monitor for SEP events of sufficiently high energies (${\geq}100$~MeV). In this study, we focus on the high-energy particle hits on Spitzer's main science instrument, IRAC, which result in saturated pixels in the infrared images taken by the camera. Such affected pixels are flagged as `radhits' and masked, in order to ensure that they are excluded from composite images of the observations. On average, IRAC measures about 4--6~radhits per second (depending on the phase of the solar cycle) due to galactic cosmic rays \citep{lowrance2018}. During 2012~May, this `baseline number' was estimated to be $4\pm0.8$~radhits per second. Thus, a number of detected radhits higher than 4 may correspond to a space weather event of solar origin.

\begin{figure}[ht]
\centering
\includegraphics[width=.55\linewidth]{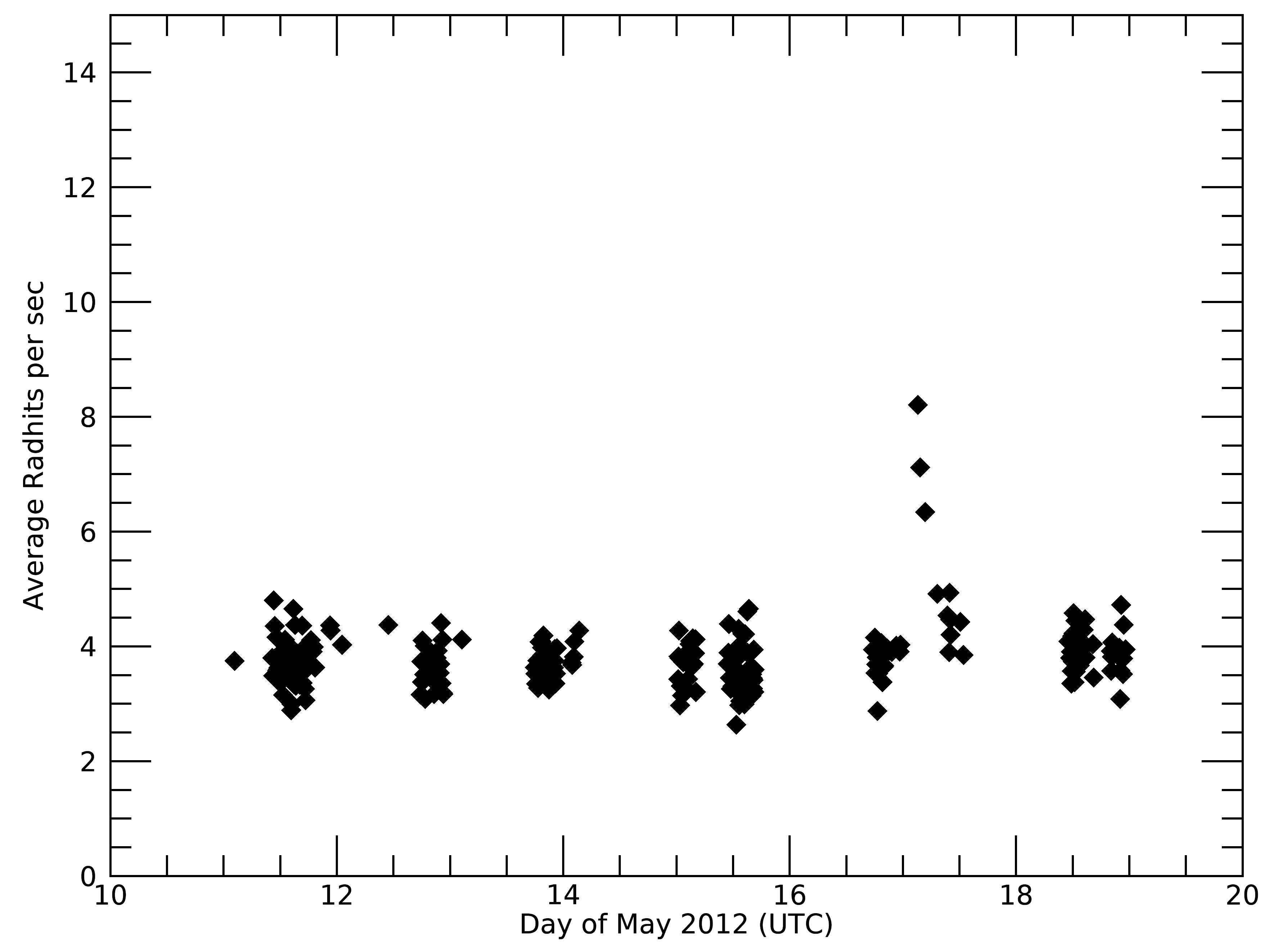}
\caption{Radhits per second measured by the IRAC instrument onboard Spitzer during the days following the 2012~May~11 CME eruption. The gaps correspond to times in which IRAC was not taking data due to downlinking. Each data point corresponds to exposure times of 12~s, 30~s, or 100~s, and is averaged over a number of frames in the range 5--25. The overall data set has an average of $3.77\pm0.61$~radhits per second.}
\label{fig:spitzer}
\end{figure}

The radhits measured by IRAC during 2012~May~10--20 are shown in Figure~\ref{fig:spitzer}. As expected, most of the data points are clustered around the value of 4. The propagation models that we used in this work (see Section~\ref{sec:models} and Table~\ref{tab:propag}) estimated a CME impact at Spitzer to take place around 2012~May~14--15. We do not observe an increase in radhits during that period, suggesting that, if the May~11 CME did indeed impact Spitzer, it was not associated with high-energy particles \citep[as it was the case in the study performed by][on a CME that erupted on 2010~November~3]{amerstorfer2018}. This is not surprising, since the ICME appeared rather slow at both Venus (Section~\ref{subsec:venus}) and Earth (Section~\ref{subsec:earth}). Nevertheless, we do observe an increase in counts starting on 2012~May~17 at 03:10~UT, with the measured radhits reaching a value ${>}8$. As it was the case for particle measurements at Venus (Section~\ref{subsec:venus}), the increase follows a several-hours data gap, hence it is not possible to establish an exact onset time for the event. Nevertheless, the timing and descending profile of the increase strongly hint towards an earlier onset and a `true peak value' higher than 8.2~radhits per second, consistent with the SEP event being associated with the May~17 eruption. Since Spitzer was separated by ${\sim}150^{\circ}$ in longitude from the flaring site, it follows that there is no possibility for the two heliolongitudes to be magnetically connected under nominal Parker spiral conditions. Hence, it is likely that the magnetic connectivity required for the observed impulsive feature of the SEPs was provided by the May~11 CME, which was being crossed by Spitzer at the time of the flare.

\subsection{Measurements at the Mars Science Laboratory} \label{subsec:msl}

At the time of the CME under study, the MSL spacecraft was approaching the end of its cruise phase, before safely landing the Curiosity rover on Mars on 2012~August~6. It was at a heliocentric distance of ${\sim}1.4$~AU and its longitude was ${\sim}35^{\circ}$ east of Earth (see Figure~\ref{fig:map}). Even though the RAD instrument was designed to measure the particle radiation environment on the surface of Mars, it was active during most of the cruise phase and collected data of the interplanetary radiation environment for future crewed missions to Mars \citep{zeitlin2013}, detecting several Forbush decreases and SEP events \citep{guo2015}.

\begin{figure}[ht]
\centering
\includegraphics[width=.55\linewidth]{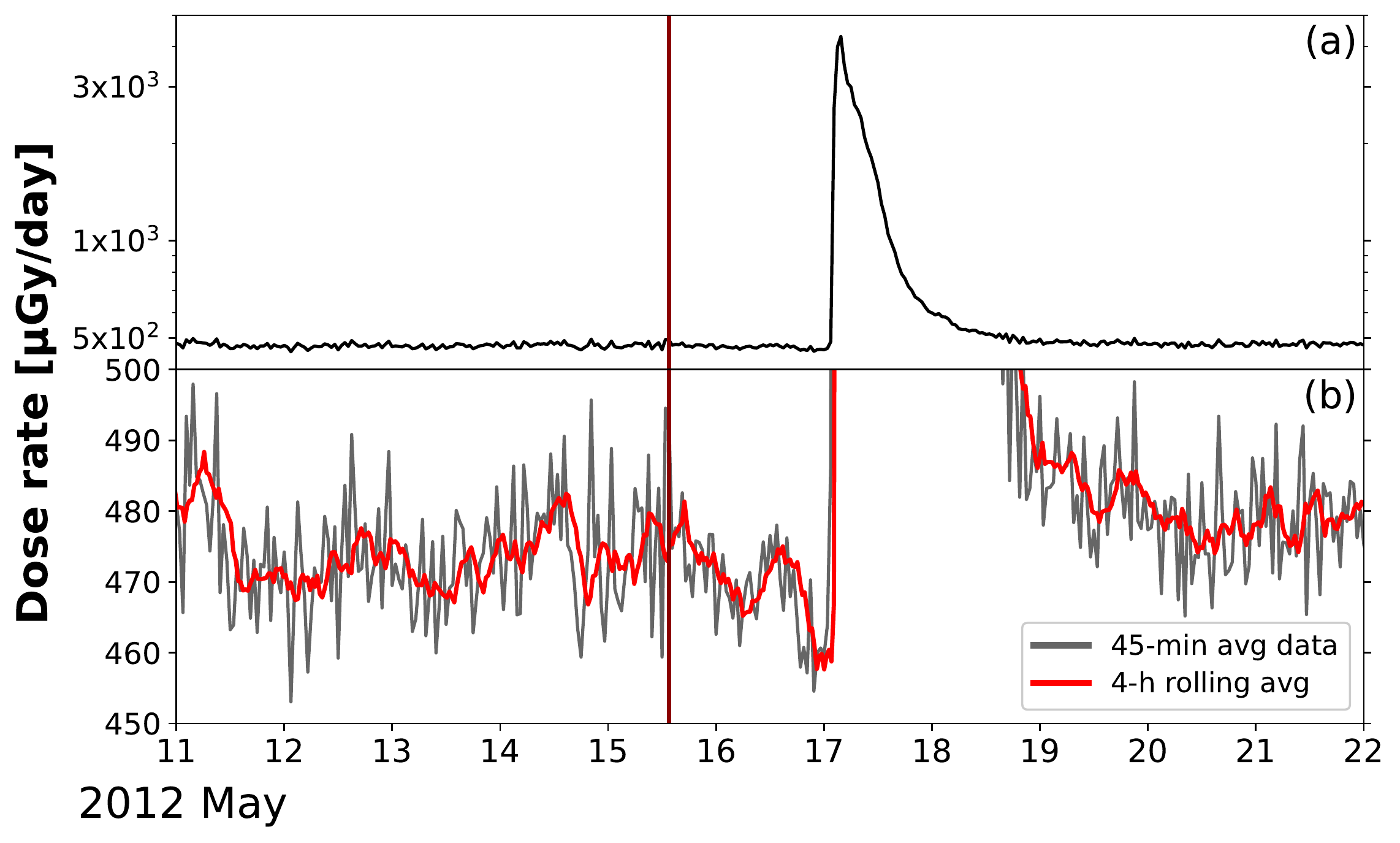}
\caption{Radiation dose measurements taken by MSL/RAD en route to Mars. Panel (b) is a zoomed-in version of the y--axis of panel (a). The time of the Forbush decrease onset (used to estimate the shock arrival time) is indicated with the solid vertical line.}
\label{fig:msl}
\end{figure}

Radiation dose measurements taken by MSL/RAD during the days following the 2012~May~11 eruption are shown in Figure~\ref{fig:msl}. Again, we find in the data clear signatures of the SEP event related to the 2012~May~17 eruption \citep[also reported by][]{battarbee2018}, starting at 02:04~UT and suggesting that the May~11 CME was also being crossed by MSL at that time. Therefore, we search for a possible Forbush decrease onset before the arrival of the SEPs. We tentatively identify such onset to take place on 2012~May~15 at 13:30~UT (marked with a solid vertical line in Figure~\ref{fig:msl}). Unfortunately, the SEP event of May~17 does not allow us to follow the full development of the Forbush decrease, hence it is not possible to declare with certainty whether the decrease is a ``classical'' two-step one. A Forbush decrease that develops in two steps usually indicates the arrival of an interplanetary shock, corresponding to the first step, and its following ICME ejecta, corresponding to the second step \citep{cane2000}. Nevertheless, the Forbush decrease onset time that we identified is consistent with the analysis performed by \citet{freiherrvonforstner2019}, who reported an arrival time at MSL on May~15, 12:00~UT (event 20120512\_01 in their study). The authors connected this Forbush decrease with the one measured at Earth earlier on the same day (see Section~\ref{subsec:earth}) and with the CME observed in HI1-B imagery (see Section~\ref{subsubsec:hi}), in agreement with our analysis of the same event.

\subsection{Measurements at Mars} \label{subsec:mars}

At 1.6~AU from the Sun and almost perfectly aligned with MSL (${<}1^{\circ}$ separation in longitude, see Figure~\ref{fig:map}) was Mars. There were no spacecraft equipped with a magnetometer in orbit around Mars at the time of this study, hence it is not possible to analyse the magnetic structure of the 2012~May~11 CME. Nevertheless, we take advantage of the extensive data sets that are available from two spacecraft to estimate the arrival time of the interplanetary shock and the boundaries of the following ICME ejecta. In particular, we complement the solar wind and particle data collected by MEX/ASPERA-3 and MOdy/HEND with measurements performed inside the Martian induced magnetosphere by MEX/MARSIS. Since Mars is not protected by an intrinsic magnetic field, the increased dynamic pressure accompanying periods of disturbed solar wind conditions is able to push the plasma boundaries of the system (e.g., bow shock, magnetic pileup boundary, ionopause) to lower altitudes quite efficiently \citep[e.g.,][]{lee2017,luhmann2017,morgan2014,sanchezcano2017,sanchezcano2020}. Hence, the level of compression of the Martian induced magnetosphere can help to determine whether a solar transient has impacted the planet. In this study, we focus on the altitude of the outbound magnetic pileup boundary crossings for successive orbits. Since the MEX orbit precession is minimal during the period that we investigate (11~days), we can assume that the boundary crossings should always occur at similar altitudes in the case of a static system.

\begin{figure}[ht]
\centering
\includegraphics[width=.6\linewidth]{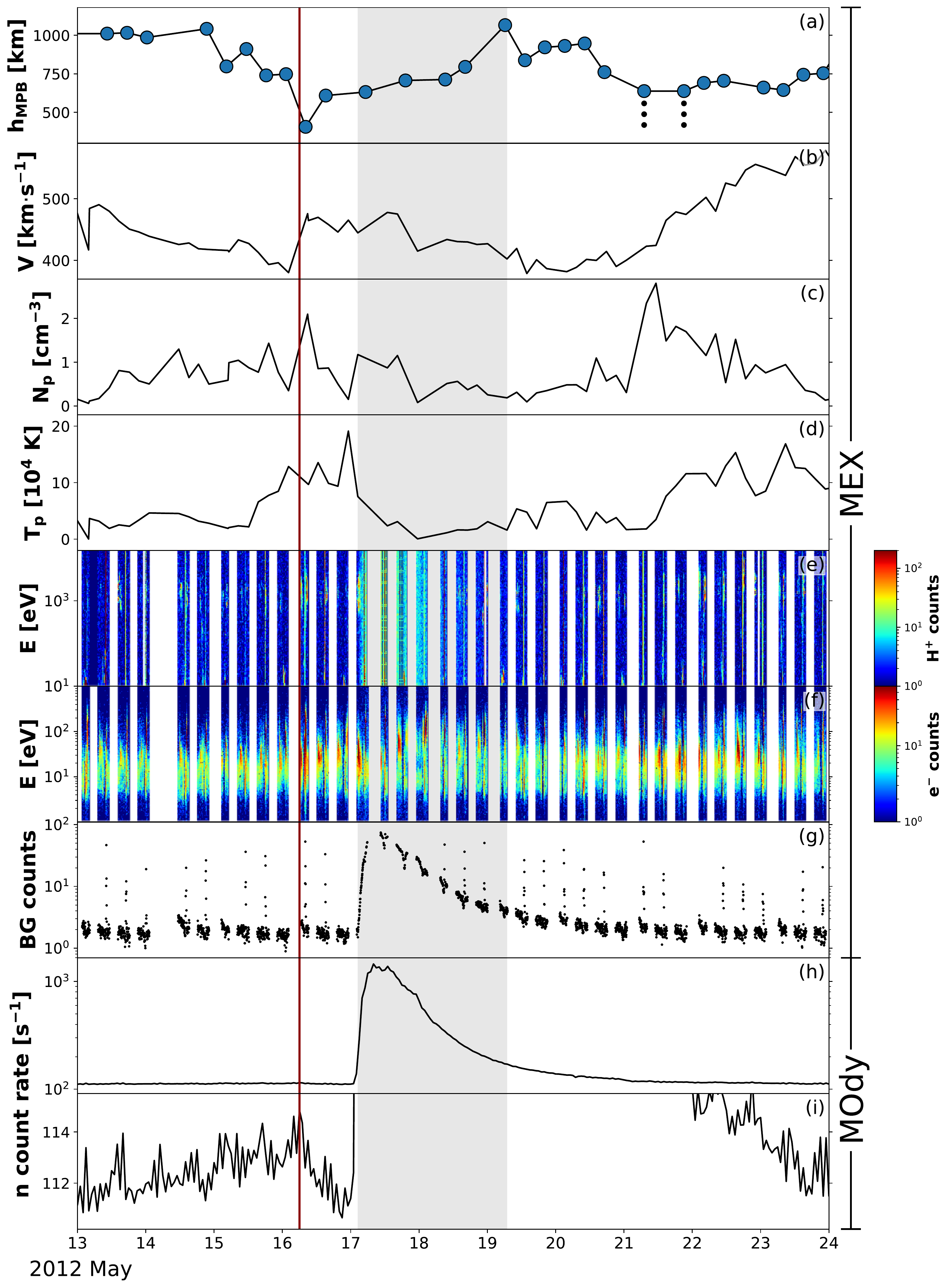}
\caption{Measurements at Mars around the expected arrival time of the 2012~May~11 CME, revealing the passage of an interplanetary disturbance. Data are taken from (a--g) MEX and (h--i) MOdy. The solid vertical line indicates the arrival of the interplanetary shock, whilst the shaded area corresponds to the estimated ejecta interval. The parameters shown are: (a) altitude of the magnetic pileup boundary (inner boundary of the magnetosheath) outbound crossings for successive orbits, (b) solar wind speed, (c) proton density, (d) proton temperature, (e) proton and (f) electron energy distribution, (g) background counts, and (h--i) neutron counts per second. The three dots under two data points in panel (a) indicate that for those orbits only an upper limit for the crossing altitude could be estimated. Panel (i) is a zoomed-in version of the y--axis of panel (h).}
\label{fig:mars}
\end{figure}

Figure~\ref{fig:mars} shows plasma and energetic particle measurements taken by MEX and MOdy following the eruption of the CME under study. Several signatures in these data indicate that an interplanetary shock (solid vertical line in Figure~\ref{fig:mars}) impacted Mars on 2012~May~16 around 06:00~UT. Namely, we observe significant compression of the Martian magnetic pileup boundary between two successive orbits (Figure~\ref{fig:mars}a, showing that the altitude of the crossings lowered from ${\sim}750$ to ${\sim}400$~km), a steep increase in the speed profile (Figure~\ref{fig:mars}b), enhancement in electron counts (Figure~\ref{fig:mars}f), and the onset of a Forbush decrease (Figure~\ref{fig:mars}i). However, as it was the case for MSL (Section~\ref{subsec:msl}), the SEP event of May~17 is seen at Mars as well, preventing us from following the whole development of the Forbush decrease and, therefore, possibly estimating the ejecta leading edge time. Nevertheless, the presence of SEP signatures, seen in both MEX/IMA background counts (Figure~\ref{fig:mars}g) and MOdy/HEND count rates (Figure~\ref{fig:mars}h) and starting at 02:16~UT, suggests that the ICME ejecta was also being crossed by Mars at the time of the May~17 eruption. If we consider the period of depressed proton temperature following the shock arrival \citep[one of the ``classic'' in-situ signatures of ejecta; e.g.,][]{richardson1995}, then the ICME ejecta boundaries would fall roughly between 2012~May~17 at 02:26~UT and 2012~May~19 at 06:59~UT (shaded region in Figure~\ref{fig:mars}). This interval is consistent with MEX/MARSIS measurements, since the estimated trailing edge coincides with an increase in the crossing altitude, suggesting that the interplanetary transient had fully travelled past Mars. These boundaries would also place the leading edge ${\sim}1$~hour after the May~17 flare onset and a few minutes after the arrival of energetic particles at Mars; however, since ASPERA-3 was not operational continuously at the time of these events, with gaps of a few hours between each observing session, the estimated boundaries are affected by large uncertainties. Nevertheless, it is reasonable to speculate that the SEP event onset and the passage of the ICME ejecta leading edge happened very close in time.

\section{Discussion} \label{sec:discussion}

In this section, we synthesise the multi-spacecraft observations, modelling results, and interpretations presented in Sections~\ref{sec:remote}, \ref{sec:models}, and \ref{sec:insitu} and discuss them in the context of three main topics: CME propagation across the inner heliosphere, CME magnetic structure, and CME role in SEP transport from the 2012~May~17 event.

\subsection{CME Propagation} \label{subsec:propagation}

The 2012~May~11 eruption was a case of a CME experiencing moderate deflection very close to the Sun, with the CME source region being located at S13E13 on the disc (Section~\ref{subsubsec:disc}) and the CME propagation direction being S10E30 from GCS reconstructions in the corona (Section~\ref{subsubsec:coronagraph}). This is not surprising, since most of the deflection is expected to take place below 30\,$R_{\odot}$ from the Sun \citep{isavnin2014}. CME deflections usually occur due to magnetic forces acting in the corona, which are dominant below 10\,$R_{\odot}$ \citep[e.g.,][]{kay2015b} and tend to divert CMEs towards the heliospheric current sheet and away from coronal holes \citep[e.g.,][]{cremades2006,kilpua2009b,xie2009}. In the case of the CME under study, the observed deflection can be likely attributed to two complementary factors, namely the global magnetic structure of the corona and the interchange reconnection scenario described in Section~\ref{subsubsec:disc}. Specifically, the CME source region was located at the edge between two highly inclined helmet streamers (from potential-field reconstructions, see Figure~S4), and reconnection of the filament's eastern leg with the nearby open field resulted in the CME diverting towards the reconnection region \citep[as shown in simulations by, e.g.,][]{lugaz2011,lynch2013,torok2011}. We remark that such a deflection, despite being considered moderate, has in general important implications for space weather forecasting: if considering the location of the source region only, then one would expect a rather frontal encounter at Earth. Only through coronagraph observations did we estimate a flank encounter at Earth and a more central one at Mars (see Figure~\ref{fig:map}a), which was also the case for the 2014~January~7 CME studied by \citet{mostl2015}. In addition, CME deflections have also implications on the structure of the IMF, thus altering the longitudinal extent that will be available for SEP acceleration and detection.

Multi-spacecraft coronagraph and HI observations permitted us to evaluate the CME propagation direction and its half-angular extent in order to estimate its impact throughout the heliosphere using different propagation models (Section~\ref{sec:models}). If we exclude modelling results from the SSEF30-A technique reported in Table~\ref{tab:propag} (which can be considered as an outlier), then we can conclude that the arrival locations predicted by the models employed in this study were consistent with each other. However, despite the agreement in terms of hit/miss, the predicted arrival times throughout the inner heliosphere featured more substantial differences. This is mostly due to the physics and assumptions involved in each model. In fact, we note that the spread in arrival times increases with both heliocentric distance and angular separation from the CME nose. In the case of approximately central encounters, the major contribution to the spread is given by the physics that regulates the CME radial propagation in each model (constant acceleration for SSE/SSSE, drag-based with constant background for DBM, and MHD-based with variable background for Enlil). In the case of flank encounters, the major contribution to the spread is instead given by the geometry of the CME front assumed in each model (spherical cross-section for SSE/SSSE, circular arc for DBM, and evolving hydrodynamical structure for Enlil). Interestingly, we note that in the case of the SSSE model the CME is predicted to reach Mars before it reaches Earth, which is due to the perfectly circular shape of the CME front. In such cases, employing a larger CME half-width may solve the issue \citep[a detailed study of how the SSSE model performs with respect to different CME half-widths is shown in][]{barnes2020}. Furthermore, we remark that \citet{palmerio2019} applied the (S)SSE model to two CMEs and noted that adding $15^{\circ}$ to the GCS-derived half-width improved arrival time predictions in both cases. Nevertheless, employing several models with different assumptions can provide an overall context useful to interpret in-situ observations, as was demonstrated in this study.

A comparison of different modelling results with the in-situ observations presented in Section~\ref{sec:insitu} is shown in Table~\ref{tab:compare}. When considering differences between the predicted and measured arrival times, it is important to remark that all the forecasts reported in Section~\ref{sec:models} were initiated using only remote-sensing data as input, and no adjustments were made to match in-situ observations. The propagation models were used as a guide to search for in-situ signatures, rather than to reproduce the observed arrivals. Most important, we remind the reader that we approximated the highly asymmetric and distorted 2012~May~11 CME with idealised, symmetrical structures in both coronagraph reconstructions (Section~\ref{subsubsec:coronagraph} and Figure~\ref{fig:gcs}) and propagation models (Section~\ref{sec:models}). Furthermore, apart from the well-known uncertainties in modelling CME arrival times that are estimated to be of the order of $\pm10$~hours at 1~AU regardless of the model used \citep[e.g.,][]{riley2018,vourlidas2019,wold2018}, additional complications may arise from the pre-existing solar wind conditions at the observing spacecraft. For example, the arrival times at Venus and Earth were likely affected by the preceding interplanetary structure (see Figures~\ref{fig:venus} and \ref{fig:earth}) that we attributed to an earlier, narrow eruption and that likely slowed down the 2012~May~11 ejecta. Furthermore, all the models predicted the impacts at MSL and Mars to take place significantly later than observed, which is possibly due to the faster wind preceding the CME at those heliolongitudes (see Figure~\ref{fig:mars}). Taking all the aforementioned factors into account, these results suggest that, even in the case of particularly complex events, models that simplify their geometry could be used to satisfactorily estimate at least the impact location(s), albeit with sometimes significant errors in arrival times.

\begin{table}[ht]
\caption{CME arrival times from three modelling techniques presented in Section~\ref{sec:models} compared with the in-situ observations presented in Section~\ref{sec:insitu}.}
\label{tab:compare}
\center
\begin{tabular}{l @{\hskip .3in} c @{\hskip .1in} c @{\hskip .1in} c @{\hskip .1in} c @{\hskip .1in} c}
\toprule
Model & Venus & Earth & Spitzer & MSL & Mars \\
\midrule
SSSE &  05/14 19:15 & 05/17 18:51 & 05/15 15:01 & 05/17 00:13 & 05/17 14:57\\
DBM & 05/14 00:08 & 05/15 00:20 & 05/14 23:42 & 05/16 14:42 & 05/17 05:49 \\
Enlil (S)& 05/13 19:33 & 05/14 21:30 & 05/14 20:03 & 05/16 05:54 & 05/16 21:58\\
Enlil (E)& 05/14 01:00 & 05/15 17:52 & 05/15 21:07 & 05/17 03:06 & 05/17 23:27\\
Observed (S) & 05/13 17:10 & 05/15 01:28 & ? & 05/15 13:30 & 05/16 06:00\\
Observed (E) & 05/14 19:23 & 05/16 15:57 & ? & ? & 05/17 02:26\\
\bottomrule
\end{tabular}
\begin{tablenotes}
\textit{Note.} Dates are shown in the format MM/DD HH:MM. The notations `(S)' and `(E)' denote shock and ejecta, respectively. The question marks indicate that it is not possible to determine whether there has been an impact, and/or at what time.
\end{tablenotes}
\end{table}

\subsection{CME Magnetic Structure} \label{subsec:mag}

The magnetic structure of the 2012~May~11 CME was inferred at several locations: at the Sun, through the solar corona, at Venus, and at Earth. The corresponding flux rope type was found to change dramatically across the different observation points. As shown in Section~\ref{subsubsec:disc}, the eruption of the 2012~May~11 CME involved the presence of a filament that disconnected asymmetrically from the Sun. The western leg stayed anchored to the photosphere for longer than the eastern one, which detached rapidly and resulted in a significant clockwise rotation of the filament. Based on these observations and on 3D reconstructions of the CME in the low corona (Figure~\ref{fig:cor1}), we estimated that the corresponding flux rope erupted as an ESW type, but was a NES type close to the Sun and a WNE type in the outer corona. The scenario of a filament eruption where one leg disconnects from the Sun early in the process whilst the other follows later was also observed by \citet{vourlidas2011}. The CME analysed in their work erupted on 2010~June~16, featured negligible rotation below 3\,$R_{\odot}$, but was observed in coronagraph imagery to rotate at an exceptionally fast rate, i.e.\ 60$^{\circ}$/day. In contrast, the 2012~May~11 CME studied here appeared to rotate significantly (${\sim}65^{\circ}$) already in the low corona. Both of these cases, however, feature the same outcome, i.e.\ that the resulting magnetic configuration in the outer corona is significantly different from that at the Sun. These rapidly rotating events are particularly challenging for space weather forecasting of CME magnetic fields \citep[e.g.,][]{kilpua2019a}, since information on the intrinsic flux rope type (inferred at the Sun) becomes practically obsolete.

Furthermore, the observed disconnection of one filament leg during the eruption raised questions about the evolution of the connectivity of the large-scale CME and whether the corresponding flux rope leg remained attached to the Sun. The relationship between the structure and evolution of erupting filaments and their overlying flux ropes is not always straightforward, and hence is an active area of research \citep[e.g.,][]{gibson2006b,howard2017,schmieder2002}. In the case of the event presented in this work, two main outcomes are possible: 1) the flux rope undergoes interchange reconnection together with the filament, hence completely detaching its western leg from the Sun, or 2) the large-scale flux rope (partially or in its entirety) maintains its field lines connected to the Sun, with the filament dynamics occurring at its periphery. We could not determine the connectivity of the flux rope from remote-sensing images alone, but measurements at Earth (Section~\ref{subsec:earth}) indicated the presence of bidirectional electrons during the first half of the ICME ejecta passage, suggesting that the front portion of the corresponding flux rope was still attached to the Sun at both ends, whilst in the rear part at least one leg was disconnected. A similar PAD profile was reported by \citet{nieveschinchilla2020}, who analysed a streamer-blowout flux rope that was observed by the Parker Solar Probe spacecraft on 2018~November~11--12. The authors provided three possible explanations for such a scenario: 1) a single (albeit distorted) flux rope transient, 2) a double flux rope, or 3) a combination of a flux rope and open magnetic field lines. Given the strong evidence for interchange reconnection presented in this work, known to be a common CME process \citep[e.g.,][]{crooker2002}, we conclude that the latter option seems the most likely.

As we remarked in Section~\ref{subsubsec:coronagraph}, the 2012~May~11 CME appeared significantly asymmetric and distorted in coronagraph imagery, especially from the SOHO and STEREO-B perspectives. Even though we approximated the CME morphology with a perfectly symmetrical GCS shell to derive its geometric and kinetic parameters (see Figure~\ref{fig:gcs}), it is important to keep in mind that the underlying flux rope may be considerably warped and that such deformations may be preserved or even enhanced in interplanetary space. Examples of deformed CME (and shock) fronts were reported by \citet{farrugia2011}, who found distortions and rotations in a magnetic cloud measured during 2007~November~19--21 by three spacecraft at 1~AU covering $40^{\circ}$ in longitude, and \citet{mostl2012}, who found inconsistent flux rope inclinations with respect to the ecliptic plane in a series of CMEs launched on 2010~August~1 and measured at various locations in the inner heliosphere covering a $120^{\circ}$ longitudinal span. From a space weather forecasting perspective, this means that knowledge of the global CME orientation at the Sun may have little to no correlation with the portion of CME that will be encountered in situ. More generally, \citet{mostl2012} suggested that the orientation of a flux rope may be viewed as a local parameter, rather than a global one. It follows that it is especially difficult to distinguish between global rotations and local deformations of a flux rope, as was pointed out by \citet{palmerio2018} who compared the orientations at the Sun with those at Earth for 20 CME events.

\begin{figure}[ht!]
\centering
\includegraphics[width=.99\linewidth]{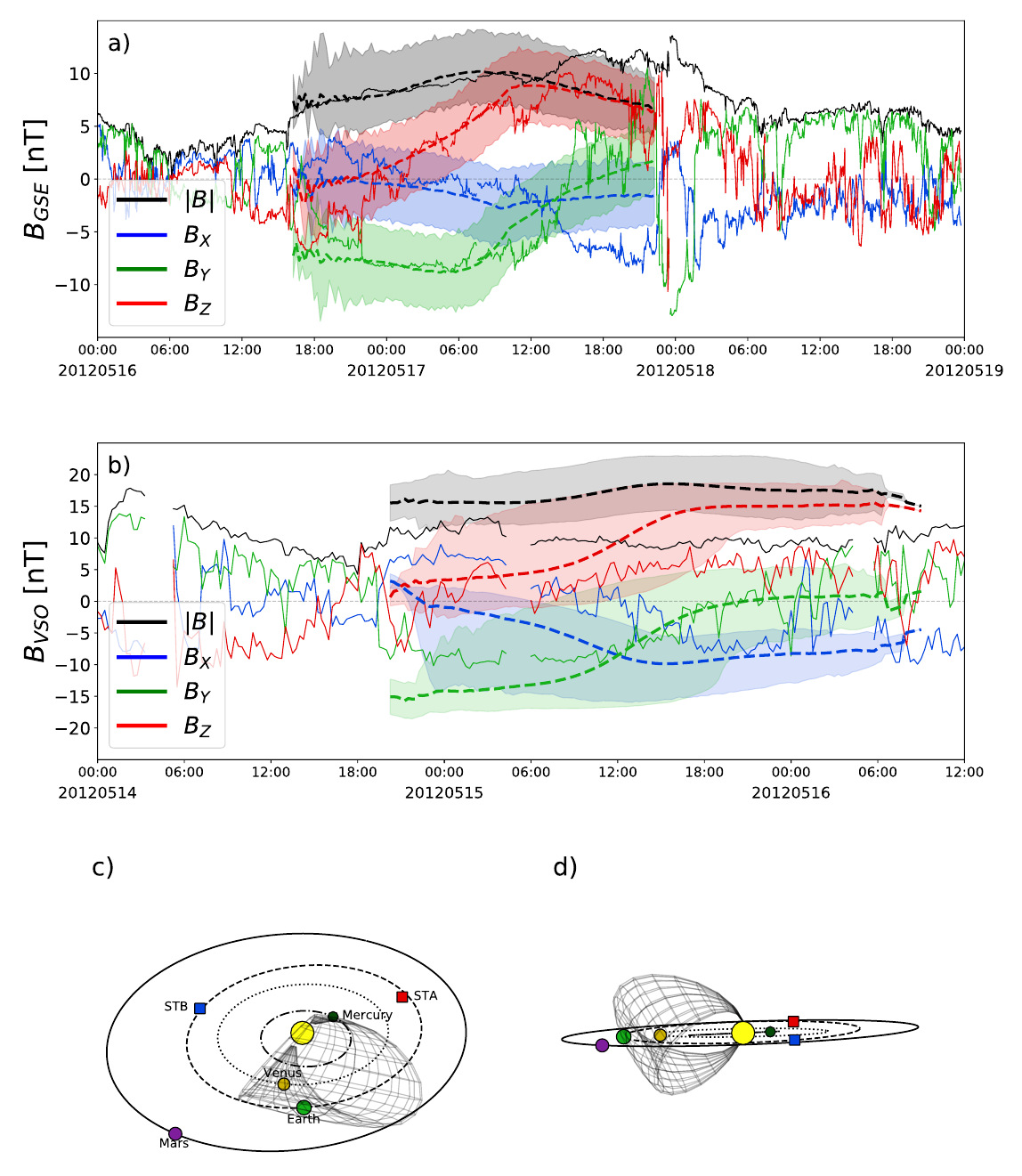}
\caption{Modelling results from the 3DCORE analysis of the 2012~May~11 CME. (a) 3DCORE ensemble fit of the ICME ejecta at Earth using measurements from Wind. (b) The fit shown in panel (a) back-propagated to Venus and superposed on VEX measurements. (c--d) Visualisation of the CME structure resulting from 3DCORE for a representative ensemble sample.}
\label{fig:3dcore}
\end{figure}

In interplanetary space, the magnetic structure of the 2012~May~11 CME could be evaluated at Venus (Section~\ref{subsec:venus}) and Earth (Section~\ref{subsec:earth}). The flux rope type and axis orientation resulted quite compatible when determined separately at the two locations but, in light of the aspects considered above, it is useful to evaluate whether the same holds true when regarding the ejecta as a coherent, rigid structure. In order to investigate this, we use the 3D Coronal Rope Ejection \citep[3DCORE;][]{mostl2018, weiss2021} modelling technique. 3DCORE is a forward simulation model that describes the structure of a CME using a torus-like geometry that is attached to the Sun and expands self-similarly as it propagates throughout the heliosphere. The expanding structure contains an embedded magnetic field that is based on an approximate analytical solution for torii \citep{vandas2017} that is similar to a Gold--Hoyle \citep{gold1960,farrugia1999} field. The fitting is performed using an approximate Bayesian computation sequential Monte Carlo algorithm, the implementation of which is described in detail in \citet{weiss2021}, that generates an ensemble of solutions. One significant advantage of this approach is that it is possible to estimate the errors on our parameters even if only using a single measurement. We initially proceed by applying the 3DCORE fitting algorithm to Wind measurements from Earth's L1 point, using boundary conditions that are very similar to those shown in Figure~\ref{fig:earth}. The fit is evaluated on the interval spanning 2012~May~17, 00:00~UT to 18:00~UT, using seven equidistant fitting points and a root-mean-square error metric. The overall time period in which ensemble solutions are accepted is set to 2012~May~16, 16:00~UT, until 2012~May~17, 22:30 UT. Figure~\ref{fig:3dcore}a shows the reproduced flux rope signatures from the ensemble at Earth and the corresponding 2-$\sigma$ spread in the magnetic field generated by the underlying uncertainties. We then back-propagate the obtained ensemble solution to Venus, allowing cross-verification of the Wind fit with the VEX measurements. This is achieved by simply re-running the simulations with the 3DCORE solution ensemble and moving the observer to Venus. The resulting magnetic field measurements from the ensemble solution at Venus, together with the propagated uncertainties, are shown in Figure~\ref{fig:3dcore}b. Finally, we also show the 3D model structure of a representative ensemble sample from two different viewing angles (Figure~\ref{fig:3dcore}c--d).

The results of the 3DCORE analysis suggest that the flux rope was oriented with a high inclination, up to $60^{\circ} \pm 10^{\circ}$, which is slightly larger than the result from the previous analysis (Sections~\ref{subsec:venus} and \ref{subsec:earth}), but nevertheless consistent with a WNE-to-SWN flux rope. Furthermore, the propagation direction of the CME is inferred to be on the opposite side of Earth when compared to the CME propagation results shown in Section~\ref{sec:models} (see, e.g., Figure~\ref{fig:enlil} for the Enlil simulation). In particular, these results predict a close miss at Mars, which is most likely not the case as there are in-situ measurements from MSL and the spacecraft orbiting Mars that strongly hint towards the contrary. These types of disagreements, however, are not out of the ordinary, as we have only fitted the in-situ magnetic field measurements at Earth's L1 point and have not added any additional constraints from other positions. Furthermore, these discrepancies may be related to CME distortion during transit and to how coherent the internal MHD structure of CMEs remains beyond ${\sim}0.3$~AU from the Sun \citep{owens2017}.

Figure~\ref{fig:3dcore}a shows that we are able to largely reconstruct the measured magnetic field profile using the 3DCORE model. The only larger discrepancy can be found in the $B_X$ component towards the end of the flux rope. Further assessment on the quality of this fit can be obtained by cross-verifying our results with measurements at Venus. In Figure~\ref{fig:3dcore}b, for simplicity, we only show the mean back-propagated ensemble of our solution at the position of the VEX spacecraft. We can conclude that, in general, the back-propagated and measured flux ropes are more or less in qualitative agreement. There is a big difference with respect to the total magnetic field strength, which can be attributed to the scaling relations that are implemented in the model. When compensating for this, the only large remaining discrepancy is again the $B_X$ component. As the VSO coordinate system is essentially the GSE equivalent for Venus, and at the time point of interest Venus and Earth are almost radially aligned, this can be expected from the same $B_X$ discrepancy at Earth. This shows that, at least to the first order, the measurements at Venus and Earth are consistent with each other and do not show any drastic evolution, clearly demonstrating that the same ICME flux rope was observed at both planets.

In conclusion, the direction of the 2012~May~11 flux rope axis was found to rotate by at least ${\sim}180^{\circ}$ clockwise between the Sun and Venus, highlighting the difficulties for space weather forecasting of the $B_{Z}$ component for rapidly rotating events. In particular, at least ${\sim}55^{\circ}$ of this rotation seems to have taken place between the last coronagraph observation and Venus, in agreement with \citet{isavnin2014} who reported that a significant amount of CME deflection and rotation can still happen between 30\,$R_{\odot}$ and 1~AU. However, we remark that, in light of the discussion above, CME rotation might have been extreme at the longitudes of Venus and Earth but not at other locations, e.g.\ at Mars, where the flux rope type could not be determined. In this regard, it is worth mentioning that the rotating CME analysed in the solar corona by \citet{vourlidas2011} was then analysed from the Sun to Earth by \citet{nieveschinchilla2012}, passing by Mercury that was separated by ${\sim}20^{\circ}$ from the Sun--Earth line. The authors found significant discrepancies in the flux rope tilt between observations in the solar corona and those at Mercury (by ${\sim}100^{\circ}$) and then from Mercury to Earth (between $20^{\circ}$ and $50^{\circ}$ depending on the chosen flux rope boundaries). The results presented in \citet{nieveschinchilla2012} and in this work show the importance of having magnetic field measurements of the same ICME at widely longitudinally separated spacecraft, in order to discern between global rotations and local distortions.

\subsection{CME Role in SEP Transport} \label{subsec:sep}

\begin{figure}[p]
\centering
\includegraphics[width=.9\linewidth]{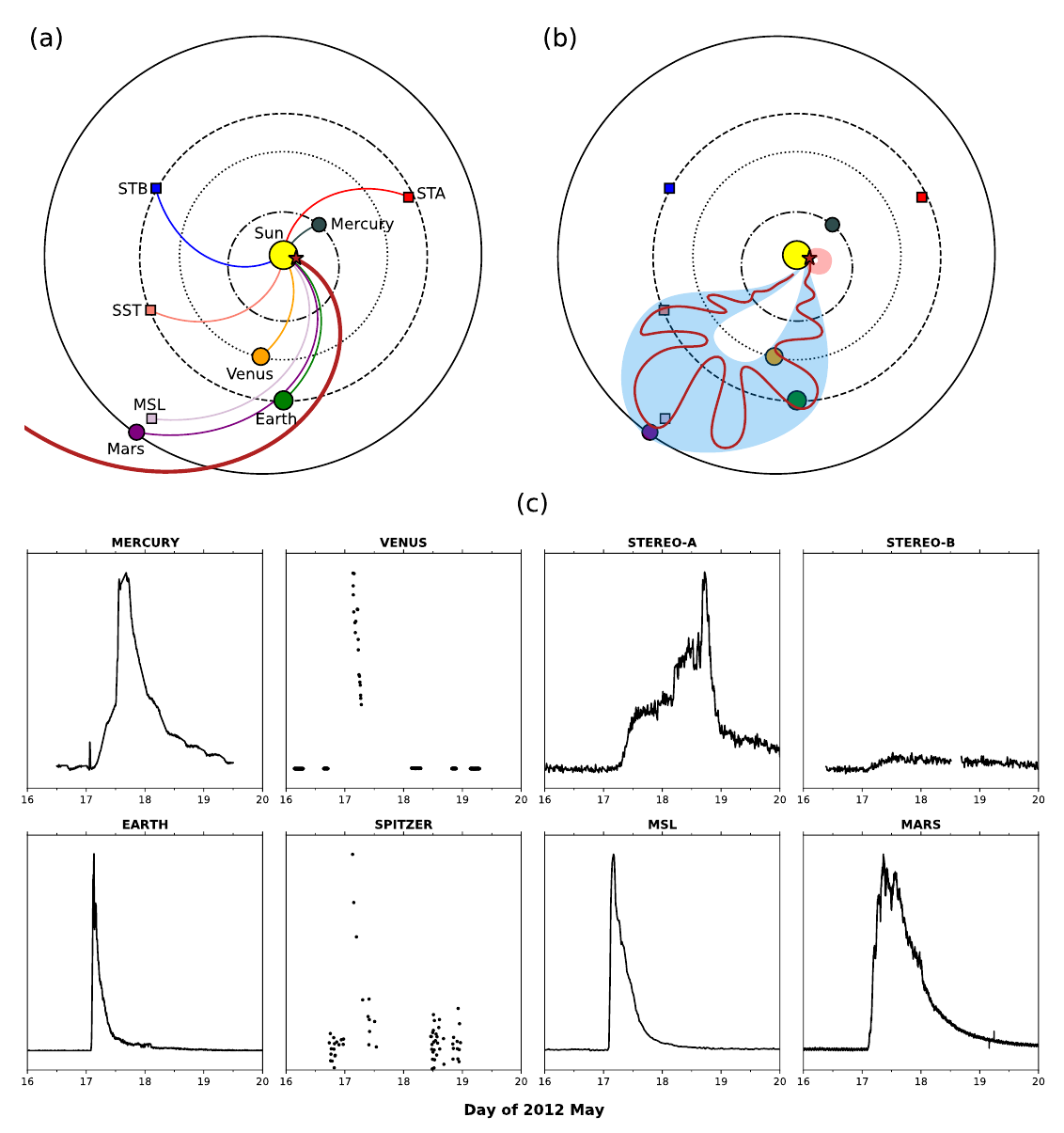}
\caption{The 2012~May~17 SEP event throughout the inner heliosphere. (a--b) Schematic to illustrate the scenario of the 2012~May~11 CME and the 2012~May~17 SEP event in the inner heliosphere. The flaring site (N11W76) is indicated with a star symbol on the edge of the circle representing the Sun. The configuration of the various planets and spacecraft is the same as in Figure~\ref{fig:map}b. (a) Hypothetical SEP transport scenario in the absence of the May~11 CME. All planets and spacecraft in the inner heliosphere have been connected to the Sun through the nominal Parker spiral for a solar wind speed of 400~km$\cdot$s$^{-1}$. The same spiral is streaming from the flare location and is represented with a thicker line. (b) The SEP transport scenario in the presence of the May~11 CME (shaded in blue) and the May~17 CME (shaded in red). SEPs propagate along the field lines within the magnetic ejecta of the May~11 CME. Connectivity to Mercury and the twin STEREO spacecraft is not shown. (c) SEP profiles (in linear scale) at all the in-situ locations shown in panels (a--b). All measurements are normalised and shown in arbitrary units, apart from STEREO-B measurements that are normalised to the STEREO-A ones. The approximate minimum energies that each shown data set is sensitive to are: 15~MeV at MESSENGER, 14~MeV at STEREO, 30~MeV at GOES, 100~MeV at Spitzer \citep{cheng2014}, 35~MeV at MSL \citep{zeitlin2013}, and 0.2~MeV at MOdy \citep{sanchezcano2018}. The minimum energy range for VEX is unknown, but it can be approximated to be similar to that of MEX (20~MeV for protons), which was derived by \citet{ramstad2018} via cross-calibration with measurements from the Mars Atmosphere and Volatile Evolution (MAVEN) spacecraft.}
\label{fig:sep}
\end{figure}

One key aim of this work was to explore the effect of the 2012~May~11 CME on the observed profiles of the SEP event that originated from the 2012~May~17 eruption. SEP signatures were indeed observed at all the locations that were predicted to be encountered by the May~11 CME (Section~\ref{sec:models}), i.e.\ Venus, Earth, Spitzer, MSL, and Mars. These findings suggest that the required magnetic connectivity for the impulsive characteristics of the in-situ SEPs was provided by the May~11 ICME, which was likely crossing all the five in-situ locations (Section~\ref{sec:insitu}) where SEP signatures were detected. This was also the conclusion drawn by \citet{rouillard2016} based on in-situ measurements at Earth only. In order to further explore this hypothesis, we first check whether SEP signatures were observed at other locations in the inner heliosphere, i.e., Mercury and the two STEREO spacecraft (see Figure~\ref{fig:map}b). Measurements from these three points were analysed by \citet[][see also Figure~\ref{fig:sep}]{battarbee2018}, who reported the presence of a gradual event at all observers, hence confirming that fast-spreading, rapidly rising SEP profiles were detected only at those locations that were being encompassed by the May~11 ICME.

Figure~\ref{fig:sep}a--b illustrates the 2012~May~17 SEP transport scenario with and without the presence of the 2012~May~11 CME. The flare site and all planets and spacecraft in the inner heliosphere have been connected through Parker spiral \citep{parker1958} field lines for a solar wind speed of 400~km$\cdot$s$^{-1}$ in Figure~\ref{fig:sep}a. As this is a hypothetical situation, we have chosen the speed of 400~km$\cdot$s$^{-1}$ to be representative of typical slow solar wind conditions close to the solar equatorial plane. It is clear from the illustration that no spacecraft in the inner heliosphere would have been perfectly magnetically connected to the eruption source region under these conditions. The longitudinal separation between the flare site and the spiral line footpoints of the locations where SEPs were observed spans from ${\sim}15^{\circ}$ (Earth) to ${\sim}85^{\circ}$ (Spitzer). The best connectivity to Earth, MSL, and Mars would be achieved for solar wind speeds in the range 310--340~km$\cdot$s$^{-1}$, whilst the best connectivity to Venus and Spitzer would be achieved for speeds in the range 170--200~km$\cdot$s$^{-1}$. Realistically speaking, it seems highly unlikely for each of these locations to be simultaneously connected to the eruptive region, which is the reason why more impulsive SEP events are usually observed over a spatially narrow region compared to gradual ones. One way to provide simultaneous connectivity to all observers would be the passage of the ICME ejecta associated with the 2012~May~11 CME, illustrated in Figure~\ref{fig:sep}b. On 2012~May~17 at 01:30~UT, the source region of the May~11 CME had rotated to S13W57, hence was in the vicinity (within ${\sim}20^{\circ}$ in both latitude and longitude) of the source region of the May~17 CME (N11W76). In such a scenario, the shock driven by the May~17 CME in the low corona \citep[which was estimated by][to have formed at $1.38\,R_{\odot}$]{gopalswamy2013} would easily intersect the western leg of the May~11 CME and accelerated particles would rapidly propagate throughout the magnetic ejecta. The energisation process itself is expected to take place on particles that are already within the May~11 CME flank, but additional SEPs could be introduced into the leg via direct or indirect transport processes such as magnetic reconnection or cross-field diffusion.

Figure~\ref{fig:sep}c shows the SEP profiles from the 2012~May~17 eruption measured in situ at eight different locations in the inner heliosphere. This makes it one of the SEP events that have been most widely observed by spacecraft at different locations. As pointed out by \citet{battarbee2018}, Mercury and STEREO-A observed a slowly rising SEP profile. Both locations did also measure the passage of an ICME: the one at Mercury (between 12:10 and 15:39~UT on May~17) was reported by \citet{winslow2015}, whilst the one at STEREO-A (between May~18 at 12:43~UT and May~19 at 09:12~UT) can be found in the STEREO ICME list \citep{jian2018}. The peak speed of 840~km$\cdot$s$^{-1}$ measured at STEREO-A is consistent with the fast May~17 CME. The gradual SEP profile at Mercury and STEREO-A is followed by an additional population of energetic storm particles (ESPs) upon the arrival of the ICME-driven shock, suggesting that particles were locally accelerated at the shock as it propagated through interplanetary space. Furthermore, STEREO-B observed a weak increase in proton flux. Such increase may be related to the May~17 CME or, alternatively, may stem from other mechanisms such as drift motion, corotation, cross-field diffusion, and turbulence \citep[as suggested by][]{battarbee2018}. We point out that the large extent and high speed of the 2012~May~17 CME, together with the other possible mechanisms enumerated above, do not rule out the fact that SEPs may have been measured at all eight locations without the presence of the May~11 CME, but it is likely that at least some of them would have experienced more slowly rising profiles such as those seen at Mercury and STEREO-A. The impulsive SEP profiles observed at Venus, Earth, Spitzer, MSL, and Mars suggest that an `instantaneous' connectivity was established through the May~11 CME, which does not exclude that a connectivity may have been established via other means. For instance, a later connection to the wide-extent and outward propagating shock of the May~17 event could also be possible for observers to the west of the event source \citep[e.g.,][]{lario2017}.

One interesting aspect to consider in the framework of IMF connectivity is the path length travelled by energetic particles up to the locations encompassed by the May~11 CME. If SEPs propagate through the preceding ejecta rather than the nominal Parker spiral, then such lengths can be expected to present somewhat higher values than `normal'. In the case under study, this analysis cannot be performed for Venus and Spitzer, where SEPs were first detected after 03:00~UT on 2012~May~17 due to data gaps. Continuous measurements at Earth, MSL, and Mars, on the other hand, enabled us to estimate precise onset times for the SEP event (01:56~UT, 02:04~UT, and 02:16~UT, respectively). These times were estimated using the Poisson--CUSUM \citep{lucas1985} method, which has been widely employed in the case of SEP events \citep[e.g.,][]{huttunenheikinmaa2005,xu2020}. Assuming 1~GeV protons \citep[which can be considered an upper energy limit for this event; e.g.,][]{kuhl2015}, a particle release time of 01:39~UT \citep[consistent with the results in][]{gopalswamy2013,li2013}, and uniform acceleration efficiency, the SEP onset times at the three locations would yield path lengths of 1.79~AU at Earth, 2.63~AU at MSL, and 3.89~AU at Mars. These values are all significantly higher than the corresponding Parker spiral lengths shown in Figure~\ref{fig:sep}a (1.18~AU at Earth, 1.91~AU at MSL, and 2.25~AU at Mars), thus suggesting that connectivity was not achieved through the nominal IMF but instead via the May~11 CME ejecta. Furthermore, the path length at Earth (1.79~AU) is consistent with the one derived by \citet{rouillard2016}, i.e.\ 1.89~AU. The large difference between the path lengths at MSL (2.63~AU) and Mars (3.89~AU), when the two locations where separated by only ${\sim}0.2$~AU, may be explained by the almost simultaneous arrival at Mars of the May~17 SEP event and the May~11 ejecta leading edge, as was shown in Section~\ref{subsec:mars}. On one hand, it is possible that optimal connectivity was reached slightly later at Mars; on the other hand, Mars might have been magnetically connected to the outer (helical) flux rope fields of the May~11 CME, which in Lunquist-like \citep[e.g.,][]{lundquist1950,lepping1990} models present the highest degree of twist, hence resulting in a longer path length.

It is not unheard that SEPs can arrive at an observer from source regions that would be considered poorly to not connected. Apart from \citet{rouillard2016} who analysed the same event described here, \citet{masson2012} studied ten GLE events between 2000 and 2006 and concluded that only three of them are consistent with particle propagation under nominal Parker spiral conditions. Five of these events occurred during the passage of CME-related disturbances past the spacecraft observing the SEPs, i.e.\ sheath regions, magnetic clouds, or ejecta rear regions, suggesting that regions on the Sun that would not otherwise be magnetically connected to Earth may be temporarily connected through transient magnetic structures. Furthermore, \citet{dresing2016} reported observations of an impulsive electron event on 2013~November~7 at both STEREO spacecraft, which were separated by $68^{\circ}$ in longitude. Whilst STEREO-A was well connected with respect to the flaring region, connectivity at STEREO-B was shown to be due to the passage of a CME that erupted three days earlier from the same source region. Indeed, as mentioned in the Introduction, observations of impulsive SEPs over large longitudinal ranges are widely documented in the literature but, to our knowledge, this is the first time that SEP transport inside a preceding ICME has been observed at five well-separated locations, distributed over $70^{\circ}$ in longitude and at four heliocentric distances between 0.7 and 1.6~AU. It is important to consider the space weather implications of such a scenario: firstly, \citet{gopalswamy2013} noted that the 2012~May~17 flare size (M5.1) was rather small for having resulted in a GLE event. One possible interpretation for the unexpected acceleration of SEPs to high energies is the twin-shock and/or twin-acceleration scenario discussed by \citet{shen2013} and \citet{ding2016}. However, the observations herein discussed are also consistent with the findings of \citet{lario2014}, who demonstrated that SEP events observed within preceding ICMEs tend to show higher peak intensities than those observed in the undisturbed solar wind. Secondly, the 2012~May~11 CME was observed at Earth as a rather slow event with almost no negative $B_{Z}$, hence its geoeffectiveness was very modest (it was associated with a $\mathrm{Dst}_{\mathrm{min}} = -43$~nT). Yet, its passage enabled observations of the first GLE event of solar cycle 24, and could have possibly represented the first GLE measured on the surface of two planets had Curiosity already landed on Mars \citep[the ``record'' went later to the 2017~September~10 event studied by, e.g.,][]{guo2018,hassler2018,lee2018,zeitlin2018}.


\section{Conclusions} \label{sec:conclusions}

In this work, we have followed the eruption and evolution of the 2012~May~11 CME and its role in spreading SEPs that originated from a later eruption on 2012~May~17. After analysing the 2012~May~11 event using remote-sensing imagery of the solar disc, corona, and inner heliosphere, we have estimated its impact throughout interplanetary space using several propagation models. Then, we have searched for signatures of the CME passage at each of the five predicted impact locations (i.e., Venus, Earth, Spitzer, MSL, and Mars). Where possible, we have studied well-known properties and phenomena usually associated with ICMEs, such as the magnetic field configuration and the associated Forbush decrease. After finding nearly-simultaneous SEP signatures at all five in-situ locations, we suggested that energetic particles accelerated by the 2012~May~17 eruption could spread over a large range of heliolongitudes due to the magnetic connectivity provided by the May~11 CME.

This work highlights the importance of using data from multiple viewpoints, both from a remote-sensing and an in-situ perspective, to characterise the complex evolution of CMEs. The event under study here appeared to be significantly distorted and rapidly rotating, which may explain why the magnetic configuration of the corresponding flux rope was not consistent across all the observation points. Some of the spacecraft employed in this study are no longer operational (e.g., MESSENGER, VEX, Spitzer, and STEREO-B), but several new missions are currently available (e.g., Parker Solar Probe, Solar Orbiter, BepiColombo, and MAVEN). Coordinated observations that employ multiple spacecraft and ground facilities will be beneficial to our understanding of CME physics, from their eruption through their interplanetary journey. International efforts such as the Whole Heliosphere and Planetary Interactions (WHPI; \url{https://whpi.hao.ucar.edu}) initiative, which aims to coordinate observations and modelling of the solar--heliospheric--planetary system during solar minimum, or the planned coordinated campaigns between Parker Solar Probe and Solar Orbiter \citep{velli2020}, are the needed step forward towards a better understanding of these complex processes.

Finally, this work showed how the interplanetary impact of the SEPs associated with the 2012~May~17 eruptive flare was tremendously influenced by the presence of the preceding May~11 ICME in the inner heliosphere. The IMF connectivity to the flare/shock acceleration site is a critical component for estimating the severity and potential impact of impulsive SEPs. While these events tend to be relatively narrow in longitudinal extent, we have shown the 2012~May~17 event was seen nearly simultaneously at five separate locations (three of which were planets) separated by up to ${\sim}150^{\circ}$ in heliographic longitude from the flaring region. Therefore, an instantaneous extrapolation of the magnetic field configuration of the flare/eruption site may not be sufficient to estimate the actual spatial extent of impulsive SEP events---the temporal history of the coronal and heliospheric field structure and evolution can also play an important role in the preconditioning of the IMF necessary for extreme events.


\section*{Sources of Data}

The HELCATS catalogues are available at \url{https://www.helcats-fp7.eu}. Images and additional information on the 2012~May~12 CME are available at \url{https://www.helcats-fp7.eu/catalogues/event_page.html?id=HCME_A__20120511_01} (STEREO-A viewpoint) and \url{https://www.helcats-fp7.eu/catalogues/event_page.html?id=HCME_B__20120512_01} (STEREO-B viewpoint).
The WSA--Enlil+Cone simulation results have been provided by the Community Coordinated Modeling Center (CCMC) at NASA Goddard Space Flight Center through their public Runs on Request system (\url{http://ccmc.gsfc.nasa.gov}). The full simulation results are available at \url{https://ccmc.gsfc.nasa.gov/database_SH/Erika_Palmerio_093020_SH_1.php}.
The Richardson~\&~Cane ICME list is available at \url{http://www.srl.caltech.edu/ACE/ASC/DATA/level3/icmetable2.htm}, whilst the NASA--Wind ICME list can be found at \url{https://wind.nasa.gov/ICMEindex.php}.
Solar disc and coronagraph data from SDO, SOHO, and STEREO are openly available at the Virtual Solar Observatory (VSO; \url{https://sdac.virtualsolar.org/}). These data were processed and analysed through SunPy \citep{sunpy2015,sunpy2020}, IDL SolarSoft \citep{bentely1998}, and the ESA JHelioviewer software \citep{muller2017}. 
Level-2 processed STEREO/HI data were obtained from the UK Solar System Data Centre (UKSSDC; \url{https://www.ukssdc.ac.uk/solar/stereo/data.html}).
GOES/XRS data were retrieved from \url{https://sohoftp.nascom.nasa.gov}.
VEX and MEX data are openly available at ESA's Planetary Science Archive (\url{https://archives.esac.esa.int/psa}). These data were processed and analysed with the aid of the \texttt{irfpy} library (\url{https://irfpy.irf.se/irfpy/index.html}).
Wind data are publicly available at NASA's Coordinated Data Analysis Web (CDAWeb) database (\url{https://cdaweb.sci.gsfc.nasa.gov/index.html/}).
Energetic particle data from GOES can be accessed at \url{https://www.ngdc.noaa.gov/stp/satellite/goes/}.
NMDB data are publicly available at \url{http://www.nmdb.eu} and Dst data can be found at \url{http://wdc.kugi.kyoto-u.ac.jp/wdc/Sec3.html}.
Spitzer data are available at the NASA/IPAC Infrared Science Archive (\url{https://irsa.ipac.caltech.edu/}).
MSL data are openly available at the Planetary Plasma Interactions (PPI) Node of NASA's Planetary Data System (PDS), accessible at \url{https://pds-ppi.igpp.ucla.edu}.
MOdy and MESSENGER data are available at the Geosciences Node of the PDS, accessible at \url{https://pds-geosciences.wustl.edu/}.
STEREO/HET data were accessed at \url{http://www.srl.caltech.edu/STEREO/Public/HET_public.html}.
The STEREO ICME list can be found at \url{https://stereo-ssc.nascom.nasa.gov/pub/ins_data/impact/level3/STEREO_Level3_ICME.pdf}.


\acknowledgments
E.~P. acknowledges the Doctoral Programme in Particle Physics and Universe Sciences (PAPU) at the University of Helsinki, the Emil Aaltonen Foundation, and the NASA Living With a Star Jack Eddy Postdoctoral Fellowship Program, administered by UCAR's Cooperative Programs for the Advancement of Earth System Science (CPAESS) under award no. NNX16AK22G.
E.~K. acknowledges the SolMAG project (ERC-COG 724391) funded by the European Research Council (ERC) in the framework of the Horizon 2020 Research and Innovation Programme, Academy of Finland project SMASH 310445, and the Finnish Centre of Excellence in Research of Sustainable Space (Academy of Finland grant no. 312390).
B.~S.-C. acknowledges support through UK-STFC grant ST/S000429/1.  
A.~W. and C.~M. thank the Austrian Science Fund (FWF): P31521-N27. 
M.~M, A.~Z., and L.~R. thank the European Space Agency (ESA) and the Belgian Federal Science Policy Office (BELSPO) for their support in the framework of the PRODEX Programme.
J.~G. thanks the Strategic Priority Program of the Chinese Academy of Sciences (grant no. XDB41000000 and XDA15017300), and the CNSA pre-research Project on Civil Aerospace Technologies (grant no. D020104).
The work of L.~T. is supported by the Academy of Finland (grant no. 322544).
We thank two anonymous reviewers, whose comments and suggestions have significantly improved this article.
We acknowledge support from the European Union FP7-SPACE-2013-1 programme for the HELCATS project (grant no. 606692).
The HI instruments on STEREO were developed by a consortium that comprised the Rutherford Appleton Laboratory (UK), the University of Birmingham (UK), Centre Spatial de Li{\`e}ge (CSL, Belgium) and the Naval Research Laboratory (NRL, USA). The STEREO/SECCHI project, of which HI is a part, is an international consortium led by NRL. We recognise the support of the UK Space Agency for funding STEREO/HI operations in the UK.
The WSA model was developed by C.~N.~Arge (currently at NASA/GSFC), and the Enlil model was developed by D.~Odstrcil (currently at GMU). We thank the model developers, M.~L.~Mays, R.~Colaninno, and the CCMC staff.
We acknowledge the NMDB, founded under the European Union's FP7 programme (contract no. 213007), for providing neutron monitor data.
We thank the WDC for Geomagnetism, Kyoto, and the geomagnetic observatories for their cooperation to make the final Dst indices available.
This work is based (in part) on archival data obtained with the Spitzer Space Telescope, which was operated by the Jet Propulsion Laboratory, California Institute of Technology, under a contract with NASA. Support for this work was provided by an award issued by JPL/Caltech.
Finally, we thank the instrument teams of all the spacecraft involved in this study.

\bibliography{bibliography.bib} 

\end{document}